\documentclass{aastex63}
\usepackage[utf8]{inputenc}
\usepackage{CJK}
\usepackage{graphicx}
\usepackage{xcolor}
\usepackage{rotating}

\received{November 25,2020}
\revised{February 23, 2021}
\accepted{March 8, 2021}
\submitjournal{ApJ}
\shorttitle{RSG in M31 and M33: mass loss rate}
\shortauthors{Wang et al.}
\begin{document}
\begin{CJK*}{UTF8}{gbsn}

\title{Red supergiants in M31 and M33 II. The Mass Loss Rate}

\author[0000-0001-5197-4858]{Tianding Wang (王天丁)}
\affiliation{Department of Astronomy, Beijing Normal University, Beijing 100875, People's Republic of China}
\email{tiandingwang@mail.bnu.edu.cn}

\correspondingauthor{Biwei Jiang}
\email{bjiang@bnu.edu.cn}

\author[0000-0003-3168-2617]{Biwei Jiang (姜碧沩)}
\affiliation{Department of Astronomy, Beijing Normal University, Beijing 100875, People's Republic of China}

\author[0000-0003-1218-8699]{Yi Ren (任逸)}
\affiliation{Department of Astronomy, Beijing Normal University, Beijing 100875, People's Republic of China}

\author[0000-0001-8247-4936]{Ming Yang (杨明)}
\affiliation{IAASARS, National Observatory of Athens, Vas. Pavlou and I. Metaxa, Penteli 15236, Greece}

\author[0000-0001-9328-4302]{Jun Li (李军)}
\affiliation{Department of Astronomy, Beijing Normal University, Beijing 100875, People's Republic of China}

%\date{\today}

\begin{abstract}
Mass loss is an important activity for red supergiants (RSGs) which can influence their evolution and final fate. Previous estimations of mass loss rates (MLRs) of RSGs exhibit significant dispersion due to the difference in method and the incompleteness of sample. With the improved quality and depth of the surveys including the UKIRT/WFCAM observation in near infrared, LGGS and PS1 in optical, a rather complete sample of RSGs is identified in M31 and M33 according to their brightness and colors. For about 2000 objects in either galaxy from this ever largest sample, the MLR is derived by fitting the observational optical-to-mid infrared spectral energy distribution (SED) with the DUSTY code of a 1-D dust radiative transfer model.  The average MLR of RSGs is found to be around $2.0\times10^{-5}{\text{M}_\odot}/\text{yr}$ with a gas-to-dust ratio of 100, which yields a total contribution to the interstellar dust by RSGs of about $1.1\times10^{-3}{\text{M}_\odot}/\text{yr}$ in M31 and $6.0 \times10^{-4}{\text{M}_\odot}/\text{yr}$ in M33, a non-negligible source in comparison with evolved low-mass stars. The MLRs are divided into three types by the dust properties, i.e. amorphous silicate, amorphous carbon and optically thin, and the relations of MLR with stellar parameters, infrared flux and colors are discussed and compared with previous works for the silicate and carbon dust group respectively.
\end{abstract}

\section{Introduction}

Mass loss is a common phenomena of RSGs, evidenced by the observed infrared excess firstly in NML Cyg and VY CMa \citep{1968ApJ...154L.125J,1969ApJ...158..619H} and lately in numerous RSGs. The destiny of a RSG  depends on mass loss rate in addition to  initial mass and chemical composition. Some RSGs may stay in the RSG stage and eventually explode as hydrogen-rich Type II-P supernovae (SNe), while the others may evolve backwards to the blue end of the H-R diagram and spend some short time as yellow supergiant stars (YSGs), blue supergiant stars (BSGs) or Wolf-Rayet stars (WRs) before exploding as SNe \citep{smartt2009,humphreys2010,2013EAS....60...31E,meynet2015,beasor2018evolution}. In either case, the strong mass loss during the RSG phase would greatly influence its ultimate fate. Moreover, the mass loss of RSGs significantly contribute to the interstellar dust content, in particular in young galaxies at high redshift where the potential dust producer --- asymptotic giant branch stars (AGBs) are not yet evolved \citep{massey2005,levesque2010}.

There have been various types of methods to estimate the mass loss rate (MLR, expressed as mass loss per year ${\rm M_\odot/yr}$) of RSGs from their infrared emission that increases with the amount of dust. One type fits the spectral energy distribution (SED) by a dust radiative transfer model (see e.g. \citealt{gordon2018}). Alternatively, spectrum analysis may be used to derive mass loss rate with the shell expansion velocities determined from the spectral line, e.g. OH double-peak maser \citep{goldman2017}. The classical Reimers law \citep{reimers1975circumstellar, kr1978} was derived from high resolution spectroscopy from which stellar wind outflow velocities and circumstellar line optical depths were determined. There are some empirical relations between the mass loss rate and stellar parameters such as luminosity or effective temperature, which is usually based on the results from the SED fitting.

RSGs in the Milky Way are the nearest though suffering heavy interstellar extinction in the Galactic plane. Still, some of them are close enough to be observed in multiple bands or even spatially resolved. The calculation leads to high MLR for some famous RSGs. \citet{gordon2018} estimated the MLR  of VX Sgr, NML Cyg and S Per from infrared imaging to be $2\sim3\times 10^{-5}{\rm M_\odot/yr}$ on average. \citet{beasor2018evolution} studied the RSGs in the clusters NGC 7419 and $\chi$ Per yielding the MLR from $10^{-7}{\rm M_\odot/yr}$ to $10^{-5}{\rm M_\odot/yr}$   and found a steep increase of MLR with stellar luminosity.  For RSGs in the Large Magellanic Cloud (LMC), \citet{van2005empirical} derived the MLR to range mostly in $10^{-4} \text{ to } 10^{-5} {\rm M_\odot/yr}$ for a sample of about 15 dust-enshrouded RSGs in the LMC  by using the DUSTY code, a dust radiative transfer model \citep{ivezic1997self}. While for a large sample ($\sim$ 30,000) of AGBs and RSGs in the LMC, \citet{Riebel2012} used the GRAMS grids and obtained a dust mass loss rate mostly between $10^{-7} \text{ to } 10^{-11} {\rm M_\odot/yr}$, a much lower MLR than by \citet{van2005empirical}. \citet{gordon2016} studied the RSGs in M31 and M33 and found the massive RSGs have an average mass loss rate of $10^{-4}{\rm M_\odot/yr}$ to $10^{-5}{\rm M_\odot/yr}$. \citet{Groenewegen2018} estimated a mass loss rate between $10^{-4} \text{ to } 10^{-10} {\rm M_\odot/yr}$ of RSGs and AGBs in LMC, SMC and dwarf spheroidal galaxies in the Local Group.

The large range of MLR in previous studies (from about $10^{-11}{\rm M_\odot/yr}$ to $10^{-4}{\rm M_\odot/yr}$) originated from the distinction  in the RSGs sample, the dust properties, the models and so on. In common, previous studies selected only a small portion of RSGs, mostly bright in optical and/or infrared, which would incline to the high MLR cases and cannot reflect the overall situation of MLR of RSGs unbiasedly. The difference in method also brings about the difficulty in comparing the results mutually. \citet[hereafter Paper I]{ren2020} constructed a sample of RSGs in M31 and M33 based mainly on the near-infrared color-color diagram and the Gaia astrometric information, which is claimed to be complete. This provides us the opportunity of investigating the MLR of all the RSGs in an individual galaxy. Besides, both M31 and M33 are spiral galaxies with similar metallicity to our Milky Way galaxy whose entire scenario cannot be viewed from internal, this would shed light on the MLR, the properties of RSGs dust and their contribution to ISM in our Galaxy. This work intends to  calculate the MLR for a couple of thousands of RSGs in M31 and M33 in this newly selected sample the majority of which is identified as a RSG for the first time, and to discuss the circumstellar dust properties and the contribution to interstellar dust by RSGs.

The paper is organized as follows. Section 2 describes the sample, and Section 3 presents the dust model we use. The last section is about the result and some discussion on MLRs of RSGs.

\section{Sample}

\subsection{Preliminary sample}

The initial sample of RSGs in M31 and M33 is taken from Paper I. \citet{lggs1,lggs2} selected 437 and 749 RSG candidates in M33 and M31 respectively by optical spectrum and colors. In comparison with the number ($\sim$ 1400) of RSGs in SMC identified by \citet{ym2018,ym2019}, the optically selected sample in M31 and M33 is far from complete. Paper I uses the deep UKIRT/WFCAM photometry in the $JHK$ bands and an innovative method to exclude the foreground stars, i.e. to remove the foreground dwarf stars from their obvious branch in the $J-H/H-K$ diagram due to their high surface gravity effect on the $H$ band. Afterwards, RSGs are recognized by their large brightness and low effective temperature, in practice from their location in the $J-K/K$ diagram that coincides well with the theoretical evolutionary track of a massive star. Consequently, Paper I identified 5498 and 3055 RSGs in M31 (along with M32 and M110) and M33 respectively, several times larger than the previous samples. After dropping off the sources outside the areas of M31 and M110,  there are 5253 and 3001 RSGs left which forms the initial sample of this work. The details of the selection and discussion on the completeness of the sample is referred to Paper I.

\subsection{Final sample: combination of catalogs in various wavebands}\label{2.2}

The initial sample is further reduced to be appropriate for estimating MLR. In complying with the traditional method, we choose to fit the SED by the dust radiative transfer model. In order to characterize the SED of a RSG, it is better to have the photometric results covering wider wavebands (in particular in the infrared) with higher accuracy. In principle, the optical bands can constrain the properties of the central star while the infrared bands demonstrate the radiation of the dust. The optical photometric data are taken from the Pan-STARRS1 large-scale survey  in the $g$, $r$, $i$, $z$ and $y$ bands \citep{chambers2016panstarrs1} and the dedicated LGGS survey in the $U$, $B$, $V$, $R$, $I$ bands \citep{lggs1, lggs2}. In addition to the UKIRT/WFCAM photometry in the $J$, $H$ and $K$ bands available for all the objects, the photometry at relatively longer wavelengths is taken from the \emph{Spitzer}/IRAC (IRAC1, IRAC2, IRAC3 and IRAC4) \citep{spitzer}, \emph{Spitzer}/MIPS24 \citep{mips} and \emph{WISE} (W1, W2, W3)\citep{wise} observations. The \emph{WISE}/W4 photometry is not used because its flux is usually much higher than that of the \emph{Spitzer}/MIPS24, which may be caused by the low resolution in this band.  These data consist an SED from optical-violet (366nm) to mid-infrared (24$\mu$m) that covers the silicate feature around 9.7$\mu$m and 18$\mu$m. Though \citet{gordon2018} reported the detection of the emission at \emph{Herschel}/70$\mu$m and 160$\mu$m for a few RSGs in the Milky Way, the photometry at such long wavelength  is unavailable for RSGs in M31 and M33 at a distance modulus of as large as $\sim$24.5 mag. Fortunately, the very high luminosity of RSGs ensures that the circumstellar dust to be hot or warm so that the lack of longer wavelength photometry has little influence on determining the MLR. On the photometric error, 0.1\,mag is set for the PS1 force PSF magnitude, LGGS and UKIRT magnitude, while the \emph{Spitzer} and \emph{WISE} data are limited to be better than 3-sigma detection. Moreover, the source is required to have eligible photometry in at least three PS1 or LGGS bands, and an IRAC or a WISE band. The final sample consists of 1741 RSGs in M31 and 1983 RSGs in M33 from the initial sample that satisfy the above conditions.

The sample used to calculate the MLR is compared with the preliminary sample in the $J-K/K$ diagram in Figure 1 for M31 and in Figure 2 for M33. The histogram in $J-K$ and $K$ exhibits no apparent systematic bias from the initial sample, and the MLR derived can then be regarded as representative of the whole sample.

\subsection{Interstellar extinction}

The interstellar extinction is subtracted before estimating the mass loss rate. For this purpose, the SFD98 \citep{sfd98} extinction map of the M31 and M33 regions is adopted. Though there are some arguments that the SFD98 map may under-estimate or over-estimate the interstellar extinction to some sightlines, it has some advantage. As the SFD98 extinction is derived from the infrared emission that comes not only from the foreground Milky Way, but also from the M31 or M33 itself, consequently the derived extinction includes both sources.  With $E(B-V)$ from SFD98, the extinction in the $U$ to $K$ band is calculated with the CCM89 law \citep{ccm1989} at $R_{\rm V}=3.1$, and the extinction in the infrared bands is calculated with the law of \citet{wang2019} and \citet{xue2016}. These laws are representative of the average Galactic extinction, but may also be plausible for M31 and M33 (see e.g. \citet{bianchi1996}) as they are spiral galaxy like the Milky Way, needless to say that the extinction law of M31 and M33 is not yet well determined. The distribution of the adopted $A_V$ for the RSGs is shown in Figure. \ref{av}. It can be seen that $A_V$ to the RSGs in M31 ranges from about 0.3 to 1.5 mag, while much smaller in M33, i.e. from 0.1-0.2 mag, and both are higher than the foreground Galactic extinction that is 0.17 mag and 0.11 mag for M31 and M33 respectively. In comparison, \citet{massey2016red} adopt a constant $A_{\rm V}=1$, which overestimates the extinction on average. It can be imagined that the over-estimation of interstellar extinction would decrease the MLR. Quantitatively, we examine the influence of interstellar extinction by running the procedure at two cases with constant $A_V=0$ and $A_V=1.0$ respectively. It is found that the mean MLR would increase by  1\% at $A_V=0$ and decrease by 5\% at $A_V=1.0$.

\subsection{Effective temperature and luminosity}

The effective temperature is calculated by  the following equation from \citet{neugent2020} with the intrinsic color index $(J-K)_0$ :
\begin{equation}\label{eq1}
    T_{\text{eff}} = 5643.5-1807.1 (J-K)_0.
\end{equation}

The luminosity of RSGs is then calculated from the $K$ band magnitude and its bolometric correction with the distance modulus of 24.32 for M31 and 24.50 for M33 (Paper I), where the bolometric correction is also taken from  \citet{neugent2020}:
\begin{equation}
    \text{BC}_K = 5.567-0.7569\times{T_{\text{eff}}}/1000.
\end{equation}

\section{The dust radiative transfer model grids}

\subsection{The model}\label{3.1}

A few dust radiative transfer models have been used to calculate the mass loss rate of RSGs. For example, \citet{van2005empirical} used the DUSTY code designed by \citet{ivezic1997self}; \citet{Riebel2012} used the GRAMS models \citep{grams}, a grid of models originally calculated by the 2-DUST code \citep{ueta2003}; \citet{Groenewegen2012} created their own dust model (MoD: More of Dust) based on DUSTY; \citet{verhoelst2009} used the MODUST model for mixed species of dust composition; \citet{liu2017} used the 2-DUST model for a 2-D dust shell.

We choose the DUSTY code for its flexibility of input parameters and its correctness has been confirmed by many applications. Because our sample of RSGs is large, we take a way similar to GRAMS that the model grids are constructed in advance, covering a wide range of dust parameters and stellar models, followed by choosing the best-match model with the observational SED. In comparison with GRAMS, our grids have higher resolution in the parameters space so that a better fitting can be obtained.

For a dust shell expanding with a constant velocity and consequently a density distribution of  $\rho{\propto}{r^{-2}}$, the MLR can be calculated from the dust and stellar parameters as following \citep{beasor2018evolution}:
\begin{equation}\label{eq2}
    \dot{M}=\frac{16\pi}{3}\frac{a}{Q_e}{\rho_d}{\phi}vr_0\tau_V
\end{equation}
where $\frac{a}{Q_e}$ refers to the ratio of the dust radius $a$ over its extinction efficiency $Q_e$, $\rho_d$ the dust density, $v$ the expansion velocity, $r_0$ the inner radius of the dust shell, and $\tau_V$ the optical depth of the dust shell at 550nm. The parameter $\phi$ refers to the gas-to-dust ratio.

The wind velocity of RSGs is difficult to measure. According to the study of AGBs and RSGs, \citet{Groenewegen2002,ramstedt2006,debeck2010,cox2012} find an expansion velocity 4 to 19 km/s. \citet {debeck2010} obtain a velocity of 14.5 and 15.4 $\text{km/s}$ for oxygen-rich and carbon-rich AGB stars respectively, which means little difference between them. In this work, we choose the \citet{goldman2017} formula for all independent of the RSG dust species (silicate or carbonaceous dust). As the the metallicity in M31 and M33 is similar to the Milky Way, the solar metallicity is used to calculate the wind speed, which leads to the wind speed as following:
\begin{equation}
    v(\text{km/s}) = 0.118(L/L_{\odot})^{0.4}.
\end{equation}

By substituting the luminosity derived above, the resultant expansion velocities vary from 4 km/s to 15 km/s, which agree with the previous works. Since the mass loss rate is proportional to the wind speed for a static wind,  it would be easy to convert for other velocity.

As the DUSTY model \citep{ivezic1997self} generates the output SED at a given luminosity $10^4{L_{\odot}}$, the MLR can be scaled as $\dot{M}=KFr_0\tau_V$ where
$F=(L/10^4L_{\odot})^{0.9}$,
$K\equiv \frac{16\pi}{3}\frac{a}{Q_e}{\rho_d}{\phi}v$ is a constant for the given luminosity,  dust species and size, ${\tau}_V$ and $r_0$ are the parameters to be derived from the model fitting.  Finally, the MLR depends on three parameters at the given dust properties: optical depth ${\tau}_V$, inner radius $r_0$ and luminosity factor $F$.

\subsection{The models library}

The model grids are built in a wide range of dust and stellar parameters as following:
\begin{enumerate}
  \item The dust species is either silicate for which the optical constants from DL84 is adopted with $\rho = 3.5 \text{g/cm}^{3}$ \citep{draine1984optical}, or amorphous carbon for which the optical constant from RM91 is adopted with $\rho = 2.26 \text{g/cm}^{3}$ \citep{rouleau1991shape}. RSGs are massive stars and originally oxygen-rich complying with cosmic abundance since no deep dredge-up like in AGB stars occurs. The silicate spectral features are already detected in many RSGs, e.g. RS Per \citep{gordon2018}, and a group of RSGs by \citet{verhoelst2009}. Both amorphous and crystalline silicate features are detected (see e.g. \citealt{liu2017}), but we only take amorphous silicate into account because there is no consensus on the composition of crystalline silicate. Moreover, the SED from multi-band photometry cannot distinguish crystalline from amorphous silicate.
  Though RSGs are supposed to be oxygen-rich, solid detections of carbonaceous dust are not rare. Early observation by \citet{sylvester1998} detected the SiC spectral feature at 11.3$\mu$m from IRC+40 427. \citet{verhoelst2009} attributed some infrared spectral features to PAHs for a few RSGs and found that amorphous carbon is needed to account for the SED of some RSGs.
  We take only amorphous carbon among the carbonaceous dust species into account because it is common in the circumstellar shell of C-rich AGB stars and the other species (PAH or SiC) is rarely visible in RSGs.
  The default optical constant of DUSTY for amorphous carbon is HN88 \citep{hanner1988}, but we change to RM91 \citep{rouleau1991shape}. This change is based on the test modeling which found that the RM91 dust can better fit the SEDs. It should be noted that the optical constants of  amorphous carbon by RM91 brings about a larger $a/Q_e$ (=0.263) than that of HN88, illustrated in   Figure \ref{qe}. If the HN88 AC were adopted as defaulted in the DUSTY code with $a/Q_e=$ 0.084, the carbon MLR would decrease by 72\%.

  \item The dust size distribution follows an MRN distribution with the maximum and minimum radius being $\max{a}=1\text{$\mu$m}, \min{a}=0.01\text{$\mu$m}$, and the power law index $q=3.5$.
  The dust size adopted in previous studies spans a wide range. \citet{van2005empirical} took one size of 0.1$\text{$\mu$m}$, and \citet{beasor2018evolution} also took one size but of 0.3$\text{$\mu$m}$.
  More studies took an MRN size distribution, including \citet{verhoelst2009,ohnaka2008,sargent2010,liu2017}, with a radius ranging from 0.01-1$\text{$\mu$m}$. \citet{gordon2018} also use the MRN size distribution, but the radius ranged from 0.005-0.25$\mu$m. To the extreme case, \citet{scicluna2015} inferred an average size of 0.5$\mu$m from their polarimetry observation of VY CMa. With only multi-band photometric data in this work, there is no possibility to discriminate these size distributions, most of them can fit the SED. Our choice of the MRN size distribution follows the classical form, and the range agrees with most of the previous studies. It should be noted that the resultant mass loss rate would depend on the size of the dust.

  \item The average ratio $a/Q_e$ is 1.573 and 0.263 for amorphous silicate and amorphous carbon respectively calculated by the following formula:
\begin{equation}
    \frac{a}{Q_{e}}=\frac{\int_{a_{min}}^{a_{max}}\frac{a^{1-q}}{Q_e(a)}da}{\int_{a_{min}}^{a_{max}}a^{-q}da}.
\end{equation}
If a constant radius is taken as done in some previous works, this parameter would change. For a constant radius of 0.1$\mu$m, this ratio changes into 0.139 and 0.052 for amorphous silicate and amorphous carbon respectively, which means the MLR would decrease by 91\% and 80\%, i.e. by an order of magnitude. Definitely this size effect should be taken into account when comparing the MLR from various works. The change of this ratio is displayed in Figure \ref{qe} for the popular values.

  \item The dust temperature at the inner radius has four options: 1200\,K, 1050\,K, 900\,K and 750\,K. The condensation temperature for silicate and amorphous carbon depends on the pressure and other factors and thus varies. \citet{gail1984formation, gail1999mineral} found that this temperature to be around 1200\,K in the circumstellar envelop around late-type star. The dust shell may be further away from the condensation line so that lower temperature is possible. Since the far-infrared data is not available, the temperature lower than 750\,K is not considered, which may under-estimate the mass loss rate, but this should not be significant for luminous RSGs.
  \item The optical depth has 100 options from 0.001 to 10 with approximately equal logarithmic interval.
  \item The stellar atmosphere model comes from the MARCS code \citep{gustafsson2008grid}.  We select the RSG models at the given mass of 15$M_\odot$, metallicity of [Z] = 0 [Z] $\equiv \log \rm{(Z/Z_{\odot})}$ where Z is the metallicity mass fraction) and surface gravity of $\log g = 0$, which are typical for a RSG in M31 and M33. Meanwhile, the effective temperature has 10 options at 3300\,K, 3400\,K, 3500\,K, 3600\,K, 3700\,K, 3800\,K, 3900\,K, 4000\,K, 4250\,K and 4500\,K. The influence of metallicity and surface gravity will be discussed later. The observational SED needs some even lower effective temperature models which is not available for massive stars, so we take a 1$M_\odot$ red-giant-like model at 2800\,K, 2900\,K, 3000\,K and 3200\,K as the substitute. This is practically feasible because the SED resolution is too low to distinguish the RSGs and RGBs spectrum at this low temperature. Moreover, only very small portion (1\%) of sources have such low temperature.
\end{enumerate}

In total, we build a library of 14 (input stellar spectrum) $\times 100$ (optical depth) $\times 4$ (inner dust temperature) $\times 2$ (dust species) = 11200 model SEDs.

\subsection{Fitting} \label{section:fitting}

The best model in the library is chosen by calculating the smallest $\chi^2$. After the flux is normalized to the $K$-band flux which is available for all sources and least affected by the dust shell among the infrared bands, the $\chi^2$ is calculated as:
\begin{equation}\label{x2}
    {\chi_i^2}={\sum^N\frac{{[f(M_i,\lambda)-f(O,\lambda)]}^2}{N|f(O,\lambda)|}}\\
    \mbox{ in which } f(M,\lambda)=\frac{F(M,\lambda)}{F(M,K)},\,\,f(O,\lambda)=\frac{F(O,\lambda)}{F(O,K)},
\end{equation}
where $F(M,\lambda)$ and $F(M,K)$ refers to the model flux in the $\lambda$ and $K$ band respectively, while  $F(O,\lambda)$ and $F(O,K)$ to the observational flux in the $\lambda$ and $K$ band respectively, and $N|f(O,\lambda)|$ is the number of observational points.

The $\chi^2$ distribution is shown in Figure \ref{chi2} after dropping 12 sources with $\chi^2 > 0.5$.
The unsuccessful cases are mainly caused by photometric error. In addition the wrong cross-identification may be responsible for a few cases, and the light variation of RSGs may bring about the disparity between optical and infrared flux.
Some RSGs have optically-thin shells with $\tau_{\mathrm V} < 0.1$ that no discrimination can be made on dust species. For such sources, the amorphous silicate is assigned according to the cosmic abundance, and they constitute an  ``optically-thin" group to be treated separately. The sources with optical depth $\tau_{\mathrm V}  \ge 0.1$  are divided into two groups by the dust species, either amorphous silicate or carbon determined by which yields the smaller $\chi^2$.

Figure \ref{spec} displays the examples for four typical cases of circumstellar dust, i.e. amorphous silicate, amorphous carbon, optically thin, and one unsuccessful case. The parameter $T_{\rm {eff}} $ derived from the model fitting is compared with that from the color index $J-K$ (Equation \ref{eq1}) in Figure \ref{teff-teff}. Because the model grids provide only some discrete values, many sources have the same $T_{\rm {eff}}$, thus the mean and the standard deviation are marked. The agreement is very well at $T_{\rm {eff}} > $3400\,K, while $T_{\rm {eff}}$ becomes smaller than that from $J-K$ at $T_{\rm {eff}} < $3400\,K. This discrepancy may be attributed to replacing the massive star model by low-mass star model at  low $T_{\rm {eff}} $, but might also be due to that Equation \ref{eq1} to transform the color index to $T_{\rm {eff}}$ is not feasible at low $T_{\rm {eff}}$.

\section{Result and discussion}

\subsection{Proportion of RSGs with silicate and carbon dust }

The results of fitting are divided into three types according to the property of circumstellar dust, (1) silicate, (2) carbon and (3) optically thin with $\tau_{\mathrm V} < 0.1$. In general, the silicate emission feature around 9.7$\mu$m leads to a relatively steep increase from the \emph{Spitzer}/IRAC3 band at $\sim5.8\mu$m to the IRAC4 band at $\sim8.0\mu$m, while the carbonaceous dust produces a plateau-like shape from 3.5$\mu$m to 8$\mu$m. For the third type, the optical depth of the circumstellar shell is very small that the infrared SED can be almost described by the stellar model, consequently there is no significant difference in taking the amorphous silicate or carbon dust. Because RSGs are supposed to be O-rich in origin, we assign Type 3 objects to have silicate dust as Type 1.

The number of objects in each type is presented in Table \ref{count}.  Among the 3712 (1733 + 1979 for M31 and M33 respectively) RSGs, 1665 (671 + 994)  sources are best fitted by silicate dust, 581 sources (329 + 252)  are attributed to amorphous carbon dust. This result means that 16\% of RSGs have carbon-rich circumstellar dust on average, with 19\% in M31 and 13\% in M33. These percentages present the statistical study of the proportion of carbon-rich dust around RSGs for the first time. It may be argued that the SED of some assigned carbon-rich RSGs can also be fitted by silicate dust. This is true if the priority is not given to the smaller $\chi^2$ but to the silicate dust. On other hand, our fitting is neither skewed to carbon dust. The infrared spectroscopy of such as JWST would be very helpful to ascertain the chemical nature. Indeed, there have been other attempts to identify the carbon dust around RSGs from photometry. For example, \citet{ym2018} attributed the difference in the mid-infrared colors related to the \emph{Spitzer} and \emph{WISE} bands to the PAH emission for the RSGs in the LMC. Basically, PAH is an indicator of carbonaceous environment. Interestingly, they found about 15\% RSGs showing the evidence of PAH, this proportion coincides with the present work. Similarly, \citet{ym2020} studied the SMC RSGs and also found the mid-infrared colors being an indicator of the PAH emission, but only 4\% being so, much smaller than in the LMC.

In comparison, the ratio of carbon-rich to oxygen-rich AGB stars is found to be 0.53 by \citet{humphreys2010} or 0.62 in the inner 4kpc for NGC 6822 which translates into an iron abundance of [Fe/H] = $-1.29$  by \citet{sibbons2012}.  It seems our derived portion of carbon-rich RSGs is not very exaggerated if compared with the AGB stars. However, the metallicity of RSGs in NGC 6822 is found to be [Z] $ \sim -0.52 $ by \citet{patrick2015}, which apparently differs from [Fe/H] = $-1.29$ derived from the C-/M-AGB ratio. RSGs and AGBs may be different populations due to the mass distinction. In spite that low- and intermediate-mass stars experience deep convection process that can dredge up the carbon element produced in helium burning to the surface to become carbon-rich,  the mechanism for a RSG to become carbon-rich is lacking though convection exists at the surface of RSGs as well. The existence of carbon-rich RSGs is still puzzling.

The proportion of carbon dust RSG is found to increase with the brightness, as shown in Figure \ref{kcp}. Interestingly the brightest RSG in both M31 and M33 has a carbon dust shell. This may be accidental because the number of bright RSGs makes no statistical significance, thus the proportion for $K < 13$ is unreliable. But the trend of increasing carbon dust with brightness continues for fainter RSG. In other words, the more massive RSGs may possess a carbon-rich dust shell more easily. In the color-color diagrams, Figure \ref{JK-K4} and \ref{JK-K8}, the carbon dust is distinguishable from the silicate dust. The RSGs with carbon dust is on average redder in $K-IRAC2$. Previous works have noticed that this color index becomes negative at low temperature for RSGs (c.f. \citet{ym2019}) which is believed due  to the CO absorption in the $IRAC2$ band. The RSG with carbon dust yet has a positive value, indicating normal continuum emission of dust, which may imply that the CO band absorption can be neglected. On the contrary, the RSG with carbon dust is bluer in $K-IRAC4$. This can be understood if the carbon-rich dust brings about some PAH emission in the $IRAC4$ band, though the dust species used in the model fitting is amorphous carbon. The discrimination of carbon and silicate dust in these two color-color diagrams generally agree with our model fitting of the dust species. Nevertheless, there does exist some confusion in both diagrams, and no clear borderline can be recognized. Definitely, the SED of some RSGs cannot be explained by silicate dust and additional dust component must be supplemented, and at present carbonaceous dust seems to the only applicable. The percentage of carbon dust derived in this work still has some uncertainty, but this comes from the model fitting. In later analysis, we will divide the sample into carbon and silicate RSGs according to the model fitting results.

\subsection{Mass Loss Rate}

The distribution of MLR is presented in Figure \ref{histmlr}. The statistical properties are listed in Table \ref{spearman}, where all the values are calculated with a gas-to-dust ratio of 100 and the column ``all" includes the optically thin sources. The MLR ranges from $10^{-9}$ to $10^{-3} {\text{M}_{\odot}} /{\rm yr}$, with a median value of $\sim 3.1\times10^{-6}{\text{M}_{\odot}} /{\rm yr}$ that is an order-of-magnitude smaller than the mean value of ${\sim} 2.0\times 10^{-5}{\text{M}_{\odot}}/{\rm yr}$, caused by numerous sources with very low MLR i.e. the optically thin sources. In general, the average MLR is about 16\% higher in M33 than in M31. On the other hand, the summed MLR of the sample RSGs in M33 ($3.97\times10^{-2}{\text{M}_{\odot}} /{\rm yr}$) is about 13\% higher than in M31 ($3.51\times10^{-2}{\text{M}_{\odot}}/{\rm yr}$).
If the RSGs are divided into two groups according to the species of the circumstellar dust, some difference appears. The MLR is higher in case of silicate dust than of carbon dust on average, and the histogram shows clearly that the carbon dust sources lack very high MLR.  In addition, the range of MLR in case of carbon dust is more concentrated. The reason may come from the difference in the dust property. As illustrated in Section \ref{3.1}, the parameter $\frac{a}{Q_e}$ is 1.573 and 0.263 for silicate and amorphous carbon respectively, to which the MLR is proportional. This implies that the MLR differs by a factor of six between silicate and carbon circumstellar shells even they have the same optical depth and inner dust radius. The density difference that $\rho_{\rm silicate} = 3.5 \text{g/cm}^{3}$ is 1.5 times $\rho_{\rm AC} = 2.26 \text{g/cm}^{3}$ yields even higher MLR in case of silicate dust. Apparently the higher-MLR RSGs with silicate dust need not be redder than those with carbon dust, which agrees with the observations.

Our results can be compared with the MLR of RSGs in previous works mentioned in Introduction. All the previous values of MLR lie apparently within the range of the present sample from $10^{-9}$ to $10^{-3} {\text{M}_{\odot}}/{\rm {yr}}$. The MLR of a few bright RSGs in the Milky Way from \citet{gordon2018} as well as the dusty RSGs in the LMC from \citet{van2005empirical} is at the high-end of our sample, which can be understood either by their high luminosity or large optical depth. The large sample of AGB and RSG stars in the LMC from \citet{Riebel2012} exhibits a range of dust MLR between $10^{-7}$ to $10^{-11}{\text{M}_\odot}/{\rm yr}$, which coincides with our results as well if the gas-to-dust ratio of 400 is adopted. Generally our results support previous works with a much more complete sample that covers various cases of MLR of RSGs.

\subsection{Relation of MLR with stellar luminosity}

The relation of MLR with stellar parameters is widely calibrated, some of which is mentioned in Introduction. With the measured $J-K$, the effective temperature $T_{\rm {eff}}$ and the bolometric correction to the $K$ magnitude are calculated in Section 2.4. The linear relation is fitted between MLR and  $\log L/{ L_{\odot}}$. Excluding the optically thin RSGs, the relations are derived for silicate and amorphous carbon dust respectively:

\begin{equation}
    \text{log}\dot{M} = 0.83{\text{log}L/L_{\odot}} -8.31 \text{\qquad  for M31 (silicate) }
\end{equation}
\begin{equation}
    \text{log}\dot{M} = 0.91{\text{log}L/L_{\odot}} -10.03 \text{\qquad  for M31 (carbon)}
\end{equation}
\begin{equation}
    \text{log}\dot{M} = 0.92{\text{log}L/L_{\odot}} -8.64 \text{\qquad  for M33 (silicate) }
\end{equation}
\begin{equation}
    \text{log}\dot{M} = 1.14{\text{log}L/L_{\odot}} -11.13 \text{\qquad  for M33 (carbon).}
\end{equation}

Table \ref{spearman} shows that the correlation is moderate.  The results are displayed in Figure \ref{mlr-lum},  and compared with previous results in Figure \ref{mlr-lum-comp}. The slope, i.e. the power law index  $\gamma$ in the relation of $\dot{M} \propto L^\gamma$, is unanimous in various cases, i.e. unanimous in  M31 and M33, and unanimous for silicate and carbon dust. This index also agrees with that from \citet{van2005empirical} and \citet{dj1988} for dusty RSGs in the LMC. It should be mentioned that, \citet{van2005empirical} and \citet{dj1988} actually derive the relations of MLR with not only luminosity but also  effective temperature, and the $\dot{M} - L $ relation displayed in Figure \ref{mlr-lum} is calculated with a mean effective temperature of our sample at 3921K. In addition, this index is around 1.0, which means that MLR is approximately proportional to luminosity. On the other hand, the index is apparently smaller than that from \citet{beasor2018evolution} for RSGs in the Galactic clusters. The discrepancy of \citet{beasor2018evolution} with others may  be caused by that their sample is different, and that they adopted a gas-to-dust ratio of 200. 

Different from the unanimous slope, the intercept, i.e. the proportional coefficient of the $\dot{M} \propto L^\gamma$ relation, is not unanimous. As discussed in Section \ref{3.1}, the amount of MLR depends on multiple factors, including the dust species, the dust size, the stellar expansion velocity and luminosity. The effects of dust species and size are discussed in previous sections, which can be as large as one order of magnitude. The expansion velocity and luminosity should have less effect, probably change the MLR by a few times. Taking all the factors into account, the MLR may differ by two orders of magnitude for a given observational SED, therefore the comparison of the vertical shift is meaningful only under the same model conditions.

Taking $T_{\text{eff}}$ into account, the relationships become:
\begin{equation}
    \text{log}\dot{M} = 0.82{\text{log}L/L_{\odot}} -0.90\text{log}T_{\text{eff}} -4.95 \text{\qquad  for M31 (silicate) }
\end{equation}
\begin{equation}
    \text{log}\dot{M} = 0.89{\text{log}L/L_{\odot}} -0.68\text{log}T_{\text{eff}} -7.49 \text{\qquad  for M31 (carbon) }
\end{equation}
\begin{equation}
    \text{log}\dot{M} = 0.73{\text{log}L/L_{\odot}} -4.84\text{log}T_{\text{eff}} -9.51 \text{\qquad  for M33 (silicate)}
\end{equation}
\begin{equation}
    \text{log}\dot{M} = 1.02{\text{log}L/L_{\odot}} -3.49\text{log}T_{\text{eff}} +1.87 \text{\qquad  for M33 (carbon). }
\end{equation}

The fitting results are displayed in Figure \ref{mlr-teff} with the value of $R^2$  which implies an improved relation in comparison with the luminosity alone. In general, the fitting is reasonably well with the residual $|\log \dot{M}_{\rm predict}-\log \dot{M}| < 1.0$, meanwhile it deviates somehow from the measurements in particular for the silicate dust. This may be understood that MLR depends on other parameters in addition to $L$ and $T_{\rm eff}$, such as stellar mass, radius or surface gravity \citep{reimers1975circumstellar,beasor2020}, and pulsation period \citep{goldman2017}. But these parameters are not conveniently available from observation.

\subsection{Relation of MLR with infrared brightness}

After examining the infrared-band brightness, a very good relation is found with the mid-infrared flux in the \emph{Spitzer}/IRAC4 band at 8$\mu$m and the MIPS24 band at 24$\mu$m. The linear fitting yields the following relations with the absolute magnitude  for which the distance modulus is 24.32 and 24.50 (Paper I) for M31 and M33 respectively.
\begin{equation}
    \text{log}\dot{M} = -0.46{M_{\rm IRAC4}} -9.37 \text{\qquad  for M31 (silicate) }
\end{equation}
\begin{equation}
    \text{log}\dot{M} = -0.36{M_{\rm IRAC4}} -9.57 \text{\qquad  for M31 (carbon) }
\end{equation}
\begin{equation}
    \text{log}\dot{M} = -0.47{M_{\rm IRAC4}} -9.58 \text{\qquad  for M33 (silicate) }
\end{equation}
\begin{equation}
    \text{log}\dot{M} = -0.42{M_{\rm IRAC4}} -10.31 \text{\qquad  for M33 (carbon) }
\end{equation}
\begin{equation}
    \text{log}\dot{M} = -0.38{M_{\rm MIPS24}} -9.48 \text{\qquad  for M31 (silicate) }
\end{equation}
\begin{equation}
    \text{log}\dot{M} = -0.37{M_{\rm MIPS24}} -10.48 \text{\qquad  for M31 (carbon) }
\end{equation}
\begin{equation}
    \text{log}\dot{M} = -0.38{M_{\rm MIPS24}} -9.68 \text{\qquad  for M33 (silicate) }
\end{equation}
\begin{equation}
    \text{log}\dot{M} = -0.43{M_{\rm MIPS24}} -11.41 \text{\qquad  for M33 (carbon) }
\end{equation}

The fitting lines are shown in Figure \ref{mlr_I4} and \ref{mlr_M1}. The correlation coefficients are all large ($< -0.7$) listed in Table \ref{spearman}, which implies a tight relation of the mid-infrared flux with the MLR. This relation is expected from the warm temperature of the circumstellar dust which emits mainly in the mid-infrared. On other hand, the relation with near-infrared flux is weak and not presented.

\subsection{Relation of MLR with infrared color indexes}

Examining the infrared colors, it is found that the MLR is sensitive to $IRAC1-IRAC4$:

\begin{equation}
    \text{log}\dot{M}=0.24 (m_{\rm IRAC1}-m_{\rm IRAC4}) -4.86 \text{\qquad  for M31 (silicate)}
\end{equation}
\begin{equation}
    \text{log}\dot{M}=0.46 (m_{\rm IRAC1}-m_{\rm IRAC4}) -6.25 \text{\qquad  for M31 (carbon)}
\end{equation}
\begin{equation}
    \text{log}\dot{M}=0.34 (m_{\rm IRAC1}-m_{\rm IRAC4}) -5.18 \text{\qquad  for M33 (silicate)}
\end{equation}
\begin{equation}
    \text{log}\dot{M}=0.64 (m_{\rm IRAC1}-m_{\rm IRAC4}) -6.50 \text{\qquad  for M33 (carbon)}
\end{equation}

As seen in Figure \ref{mlr-I1-I4}, the scattering around the fitting line is significant, and the correlation of MLR with the color index is not so tight as with the mid-infrared flux. The correlation coefficients in Table \ref{spearman} also manifest that the correlation is very weak for silicate dust, and moderate for carbon dust. Previously, \citet{Groenewegen2018} and \citet{matsuura2009} derived the relation of MLR with $IRAC1-IRAC4$ as well for AGB stars, but they take an exponential function as illustrated in Figure \ref{mlr-I1-I4} by the dash-dot lines which hardly delineate the tendency for the RSGs sample. It seems no good relation exists for the MLR with the infrared colors.

\subsection{Total mass loss rate}\label{4.6}

The initial sample from Paper I should be complete since the lower limit of the RSG brightness in M31 and M33 is higher than the observational complete magnitude. But the sample for which we derived the MLR is incomplete after removing those without appropriate photometric SED. Nevertheless, the comparison with the initial sample shown in Figure \ref{count} finds no bias to the $K$ magnitude or the $J-K$ colors. Thus we simply multiply the summed MLR from our sample by the ratio of the initial sample to the final sample, i.e.  $5253/1733=3.03$ for M31 and $3001/1979=1.52$ for M33. As listed in Table  \ref{mean}, the summed mass loss rate of all the modelled sample RSGs in M31 is $\sim0.035{\text{M}_\odot}/\text{yr}$, and in M33 is $\sim0.040{\text{M}_\odot}/\text{yr}$. After multiplying by this ratio, the total MLR of RSG is then $0.11 {\text{M}_\odot}/\text{yr}$ in M31 and $0.060 {\text{M}_\odot}/\text{yr}$ in M33. Dividing by the gas-to-dust ratio, i.e. 100 in this work, the dust production rate by RSGs is $10^{-3}{\text{M}_\odot}/\text{yr}$ in M31 and $6 \times10^{-4}{\text{M}_\odot}/\text{yr}$ in M33. At present, we have no idea on the dust mass contributed by AGB stars in M31 or M33. Taking the Milky Way galaxy as a reference, the dust production rate by evolved low mass stars is a few $10^{-3}{\text{M}_\odot}/\text{yr}$ according to \citet{2009ASPC..414..453D}, or about $10^{-3}{\text{M}_\odot}/\text{yr}$ according to \citet{jones2001}. Though the uncertainty of these values may be a few factors, it seems that RSGs can be a non-negligible  source of interstellar dust. The precise estimation of the portion of RSGs dust can be made only after a systematic calculation of the MLR of AGB stars in M31 and M33 in our further work.

\subsection{Influence of metallicity and surface gravity}

Metallicity and surface gravity change the stellar spectrum as well as effective temperature, which are assumed to be constant in modelling. The metallicity of both M31 \citep{1982ApJ...254...50B} and M33 \citep{u2009} is found to exhibit some gradient with the distance from the center. Considering that RSGs are located in the spiral arm or ring area, the case of [Z] = 0.25 with the MARCS RSGs model is examined. Although \citep{goldman2017} derived a relation of the wind speed proportional to metallictiy, a scale of [Z] = 0.25 would yield an unreasonably high expansion velocity.  Moreover,  they suggest that the mass-loss rate is independent of metallicity between a half and twice solar. Thus the velocity is still calculated with  [Z]=0. The resultant MLR is listed in Table \ref{z_logg}. It can been seen that a higher metallicity will bring about a higher MLR, which is possibly caused by a lower bump in the $J, H, K$ bands.

Quantitatively, the mean MLR shows very little difference when [Z] changes from 0 to 0.25, with a number of $2.03\times10^{-5}{\text{M}_\odot}/\text{yr}$ meaning an increase of 1\%.  For the surface gravity, the case of $\log g=-0.5$ is examined. As shown in Table \ref{z_logg}, the mean MLR increases to $2.03\times10^{-5}{\text{M}_\odot}/\text{yr}$ with $\log g=-0.5$, i.e. 1\%. To sum up, the influence of stellar metallicity and surface gravity is very small, only about one percent.

\acknowledgments We are grateful to Profs. Yong Zhang, Junichi Nakshima, Jian Gao, Aigen Li and Hai-Bo Yuan for very helpful discussions, and to  Mingxu Sun, Ye Wang and Tongtian Ren for their discussions. We thank the referee for his/her very helpful suggestion.  This work is supported by National Key R\&D Program of China No. 2019YFA0405503, the CSST Milky Way Survey on Dust and Extinction Project and NSFC 11533002. This work has made use of data from UKIRT, PS1, LGGS, \emph{WISE} and \emph{Spitzer}.

\software{Astropy \citep{2013A&A...558A..33A}, TOPCAT \citep{topcat}}

\bibliography{ref}{}

\begin{thebibliography}{}
\expandafter\ifx\csname natexlab\endcsname\relax\def\natexlab#1{#1}\fi
\providecommand{\url}[1]{\href{#1}{#1}}
\providecommand{\dodoi}[1]{doi:~\href{http://doi.org/#1}{\nolinkurl{#1}}}
\providecommand{\doeprint}[1]{\href{http://ascl.net/#1}{\nolinkurl{http://ascl.net/#1}}}
\providecommand{\doarXiv}[1]{\href{https://arxiv.org/abs/#1}{\nolinkurl{https://arxiv.org/abs/#1}}}

\bibitem[{{Astropy Collaboration} {et~al.}(2013){Astropy Collaboration},
  {Robitaille}, {Tollerud}, {Greenfield}, {Droettboom}, {Bray}, {Aldcroft},
  {Davis}, {Ginsburg}, {Price-Whelan}, {Kerzendorf}, {Conley}, {Crighton},
  {Barbary}, {Muna}, {Ferguson}, {Grollier}, {Parikh}, {Nair}, {Unther},
  {Deil}, {Woillez}, {Conseil}, {Kramer}, {Turner}, {Singer}, {Fox}, {Weaver},
  {Zabalza}, {Edwards}, {Azalee Bostroem}, {Burke}, {Casey}, {Crawford},
  {Dencheva}, {Ely}, {Jenness}, {Labrie}, {Lim}, {Pierfederici}, {Pontzen},
  {Ptak}, {Refsdal}, {Servillat}, \& {Streicher}}]{2013A&A...558A..33A}
{Astropy Collaboration}, {Robitaille}, T.~P., {Tollerud}, E.~J., {et~al.} 2013,
  \aap, 558, A33, \dodoi{10.1051/0004-6361/201322068}

\bibitem[{{Beasor} \& {Davies}(2018)}]{beasor2018evolution}
{Beasor}, E.~R., \& {Davies}, B. 2018, \mnras, 475, 55,
  \dodoi{10.1093/mnras/stx3174}

\bibitem[{{Beasor} {et~al.}(2020){Beasor}, {Davies}, {Smith}, {van Loon},
  {Gehrz}, \& {Figer}}]{beasor2020}
{Beasor}, E.~R., {Davies}, B., {Smith}, N., {et~al.} 2020, \mnras, 492, 5994,
  \dodoi{10.1093/mnras/staa255}

\bibitem[{{Bianchi} {et~al.}(1996){Bianchi}, {Clayton}, {Bohlin}, {Hutchings},
  \& {Massey}}]{bianchi1996}
{Bianchi}, L., {Clayton}, G.~C., {Bohlin}, R.~C., {Hutchings}, J.~B., \&
  {Massey}, P. 1996, \apj, 471, 203, \dodoi{10.1086/177963}

\bibitem[{{Blair} {et~al.}(1982){Blair}, {Kirshner}, \&
  {Chevalier}}]{1982ApJ...254...50B}
{Blair}, W.~P., {Kirshner}, R.~P., \& {Chevalier}, R.~A. 1982, \apj, 254, 50,
  \dodoi{10.1086/159703}

\bibitem[{{Cardelli} {et~al.}(1989){Cardelli}, {Clayton}, \&
  {Mathis}}]{ccm1989}
{Cardelli}, J.~A., {Clayton}, G.~C., \& {Mathis}, J.~S. 1989, \apj, 345, 245,
  \dodoi{10.1086/167900}

\bibitem[{Chambers {et~al.}(2016)Chambers, Magnier, Metcalfe, Flewelling,
  Huber, Waters, Denneau, Draper, Farrow, Finkbeiner, Holmberg, Koppenhoefer,
  Price, Rest, Saglia, Schlafly, Smartt, Sweeney, Wainscoat, Burgett, Chastel,
  Grav, Heasley, Hodapp, Jedicke, Kaiser, Kudritzki, Luppino, Lupton, Monet,
  Morgan, Onaka, Shiao, Stubbs, Tonry, White, Bañados, Bell, Bender, Bernard,
  Boegner, Boffi, Botticella, Calamida, Casertano, Chen, Chen, Cole, Deacon,
  Frenk, Fitzsimmons, Gezari, Gibbs, Goessl, Goggia, Gourgue, Goldman, Grant,
  Grebel, Hambly, Hasinger, Heavens, Heckman, Henderson, Henning, Holman, Hopp,
  Ip, Isani, Jackson, Keyes, Koekemoer, Kotak, Le, Liska, Long, Lucey, Liu,
  Martin, Masci, McLean, Mindel, Misra, Morganson, Murphy, Obaika, Narayan,
  Nieto-Santisteban, Norberg, Peacock, Pier, Postman, Primak, Rae, Rai, Riess,
  Riffeser, Rix, Röser, Russel, Rutz, Schilbach, Schultz, Scolnic, Strolger,
  Szalay, Seitz, Small, Smith, Soderblom, Taylor, Thomson, Taylor, Thakar,
  Thiel, Thilker, Unger, Urata, Valenti, Wagner, Walder, Walter, Watters,
  Werner, Wood-Vasey, \& Wyse}]{chambers2016panstarrs1}
Chambers, K.~C., Magnier, E.~A., Metcalfe, N., {et~al.} 2016, The Pan-STARRS1
  Surveys.
\newblock \doarXiv{1612.05560}

\bibitem[{{Cox} {et~al.}(2012){Cox}, {Kerschbaum}, {van Marle}, {Decin},
  {Ladjal}, {Mayer}, {Groenewegen}, {van Eck}, {Royer}, {Ottensamer}, {Ueta},
  {Jorissen}, {Mecina}, {Meliani}, {Luntzer}, {Blommaert}, {Posch},
  {Vandenbussche}, \& {Waelkens}}]{cox2012}
{Cox}, N.~L.~J., {Kerschbaum}, F., {van Marle}, A.~J., {et~al.} 2012, \aap,
  537, A35, \dodoi{10.1051/0004-6361/201117910}

\bibitem[{{De Beck} {et~al.}(2010){De Beck}, {Decin}, {de Koter}, {Justtanont},
  {Verhoelst}, {Kemper}, \& {Menten}}]{debeck2010}
{De Beck}, E., {Decin}, L., {de Koter}, A., {et~al.} 2010, \aap, 523, A18,
  \dodoi{10.1051/0004-6361/200913771}

\bibitem[{{de Jager} {et~al.}(1988){de Jager}, {Nieuwenhuijzen}, \& {van der
  Hucht}}]{dj1988}
{de Jager}, C., {Nieuwenhuijzen}, H., \& {van der Hucht}, K.~A. 1988, \aaps,
  72, 259

\bibitem[{{Draine}(2009)}]{2009ASPC..414..453D}
{Draine}, B.~T. 2009, in Astronomical Society of the Pacific Conference Series,
  Vol. 414, Cosmic Dust - Near and Far, ed. T.~{Henning}, E.~{Gr{\"u}n}, \&
  J.~{Steinacker}, 453.
\newblock \doarXiv{0903.1658}

\bibitem[{{Draine} \& {Lee}(1984)}]{draine1984optical}
{Draine}, B.~T., \& {Lee}, H.~M. 1984, \apj, 285, 89, \dodoi{10.1086/162480}

\bibitem[{{Ekstr{\"o}m} {et~al.}(2013){Ekstr{\"o}m}, {Georgy}, {Meynet},
  {Groh}, \& {Granada}}]{2013EAS....60...31E}
{Ekstr{\"o}m}, S., {Georgy}, C., {Meynet}, G., {Groh}, J., \& {Granada}, A.
  2013, in EAS Publications Series, Vol.~60, EAS Publications Series, ed.
  P.~{Kervella}, T.~{Le Bertre}, \& G.~{Perrin}, 31--41,
  \dodoi{10.1051/eas/1360003}

\bibitem[{{Gail} \& {Sedlmayr}(1984)}]{gail1984formation}
{Gail}, H.~P., \& {Sedlmayr}, E. 1984, \aap, 132, 163

\bibitem[{{Gail} \& {Sedlmayr}(1999)}]{gail1999mineral}
---. 1999, \aap, 347, 594

\bibitem[{{Goldman} {et~al.}(2017){Goldman}, {van Loon}, {Zijlstra}, {Green},
  {Wood}, {Nanni}, {Imai}, {Whitelock}, {Matsuura}, {Groenewegen}, \&
  {G{\'o}mez}}]{goldman2017}
{Goldman}, S.~R., {van Loon}, J.~T., {Zijlstra}, A.~A., {et~al.} 2017, \mnras,
  465, 403, \dodoi{10.1093/mnras/stw2708}

\bibitem[{{Gordon} {et~al.}(2016){Gordon}, {Humphreys}, \&
  {Jones}}]{gordon2016}
{Gordon}, M.~S., {Humphreys}, R.~M., \& {Jones}, T.~J. 2016, \apj, 825, 50,
  \dodoi{10.3847/0004-637X/825/1/50}

\bibitem[{{Gordon} {et~al.}(2018){Gordon}, {Humphreys}, {Jones}, {Shenoy},
  {Gehrz}, {Helton}, {Marengo}, {Hinz}, \& {Hoffmann}}]{gordon2018}
{Gordon}, M.~S., {Humphreys}, R.~M., {Jones}, T.~J., {et~al.} 2018, \aj, 155,
  212, \dodoi{10.3847/1538-3881/aab961}

\bibitem[{{Groenewegen}(2012)}]{Groenewegen2012}
{Groenewegen}, M.~A.~T. 2012, \aap, 543, A36,
  \dodoi{10.1051/0004-6361/201218965}

\bibitem[{{Groenewegen} {et~al.}(2002){Groenewegen}, {Sevenster}, {Spoon}, \&
  {P{\'e}rez}}]{Groenewegen2002}
{Groenewegen}, M.~A.~T., {Sevenster}, M., {Spoon}, H.~W.~W., \& {P{\'e}rez}, I.
  2002, \aap, 390, 511, \dodoi{10.1051/0004-6361:20020728}

\bibitem[{{Groenewegen} \& {Sloan}(2018)}]{Groenewegen2018}
{Groenewegen}, M.~A.~T., \& {Sloan}, G.~C. 2018, \aap, 609, A114,
  \dodoi{10.1051/0004-6361/201731089}

\bibitem[{{Gustafsson} {et~al.}(2008){Gustafsson}, {Edvardsson}, {Eriksson},
  {J{\o}rgensen}, {Nordlund}, \& {Plez}}]{gustafsson2008grid}
{Gustafsson}, B., {Edvardsson}, B., {Eriksson}, K., {et~al.} 2008, \aap, 486,
  951, \dodoi{10.1051/0004-6361:200809724}

\bibitem[{{Hanner}(1988)}]{hanner1988}
{Hanner}, M. 1988, {Grain optical properties.}, Infrared Observations of Comets
  Halley and Wilson and Properties of the Grains

\bibitem[{{Humphreys}(2010)}]{humphreys2010}
{Humphreys}, R.~M. 2010, in Astronomical Society of the Pacific Conference
  Series, Vol. 425, Hot and Cool: Bridging Gaps in Massive Star Evolution, ed.
  C.~{Leitherer}, P.~D. {Bennett}, P.~W. {Morris}, \& J.~T. {Van Loon}, 247

\bibitem[{{Hyland} {et~al.}(1969){Hyland}, {Becklin}, {Neugebauer}, \&
  {Wallerstein}}]{1969ApJ...158..619H}
{Hyland}, A.~R., {Becklin}, E.~E., {Neugebauer}, G., \& {Wallerstein}, G. 1969,
  \apj, 158, 619, \dodoi{10.1086/150224}

\bibitem[{{Ivezic} \& {Elitzur}(1997)}]{ivezic1997self}
{Ivezic}, Z., \& {Elitzur}, M. 1997, \mnras, 287, 799,
  \dodoi{10.1093/mnras/287.4.799}

\bibitem[{{Johnson}(1968)}]{1968ApJ...154L.125J}
{Johnson}, H.~L. 1968, \apjl, 154, L125, \dodoi{10.1086/180284}

\bibitem[{{Jones}(2001)}]{jones2001}
{Jones}, A.~P. 2001, Philosophical Transactions of the Royal Society of London
  Series A, 359, 1961, \dodoi{10.1098/rsta.2001.0890}

\bibitem[{{Kudritzki} \& {Reimers}(1978)}]{kr1978}
{Kudritzki}, R.~P., \& {Reimers}, D. 1978, \aap, 70, 227

\bibitem[{Levesque(2010)}]{levesque2010}
Levesque, E.~M. 2010, New Astronomy Reviews, 54, 1,
  \dodoi{https://doi.org/10.1016/j.newar.2009.10.002}

\bibitem[{{Liu} \& {Jiang}(2017)}]{liu2017}
{Liu}, J., \& {Jiang}, B. 2017, \aj, 153, 176, \dodoi{10.3847/1538-3881/aa6334}

\bibitem[{Massey \& Evans(2016)}]{massey2016red}
Massey, P., \& Evans, K.~A. 2016, The Astrophysical Journal, 826, 224

\bibitem[{{Massey} {et~al.}(2007){Massey}, {Olsen}, {Hodge}, {Jacoby},
  {McNeill}, {Smith}, \& {Strong}}]{lggs2}
{Massey}, P., {Olsen}, K.~A.~G., {Hodge}, P.~W., {et~al.} 2007, \aj, 133, 2393,
  \dodoi{10.1086/513319}

\bibitem[{{Massey} {et~al.}(2006){Massey}, {Olsen}, {Hodge}, {Strong},
  {Jacoby}, {Schlingman}, \& {Smith}}]{lggs1}
---. 2006, \aj, 131, 2478, \dodoi{10.1086/503256}

\bibitem[{{Massey} {et~al.}(2005){Massey}, {Plez}, {Levesque}, {Olsen},
  {Clayton}, \& {Josselin}}]{massey2005}
{Massey}, P., {Plez}, B., {Levesque}, E.~M., {et~al.} 2005, \apj, 634, 1286,
  \dodoi{10.1086/497065}

\bibitem[{{Matsuura} {et~al.}(2009){Matsuura}, {Barlow}, {Zijlstra},
  {Whitelock}, {Cioni}, {Groenewegen}, {Volk}, {Kemper}, {Kodama}, {Lagadec},
  {Meixner}, {Sloan}, \& {Srinivasan}}]{matsuura2009}
{Matsuura}, M., {Barlow}, M.~J., {Zijlstra}, A.~A., {et~al.} 2009, \mnras, 396,
  918, \dodoi{10.1111/j.1365-2966.2009.14743.x}

\bibitem[{{Meynet} {et~al.}(2015){Meynet}, {Chomienne}, {Ekstr{\"o}m},
  {Georgy}, {Granada}, {Groh}, {Maeder}, {Eggenberger}, {Levesque}, \&
  {Massey}}]{meynet2015}
{Meynet}, G., {Chomienne}, V., {Ekstr{\"o}m}, S., {et~al.} 2015, \aap, 575,
  A60, \dodoi{10.1051/0004-6361/201424671}

\bibitem[{{Neugent} {et~al.}(2020){Neugent}, {Massey}, {Georgy}, {Drout},
  {Mommert}, {Levesque}, {Meynet}, \& {Ekstr{\"o}m}}]{neugent2020}
{Neugent}, K.~F., {Massey}, P., {Georgy}, C., {et~al.} 2020, \apj, 889, 44,
  \dodoi{10.3847/1538-4357/ab5ba0}

\bibitem[{{Ohnaka} {et~al.}(2008){Ohnaka}, {Izumiura}, {Leinert}, {Driebe},
  {Weigelt}, \& {Wittkowski}}]{ohnaka2008}
{Ohnaka}, K., {Izumiura}, H., {Leinert}, C., {et~al.} 2008, \aap, 490, 173,
  \dodoi{10.1051/0004-6361:200810229}

\bibitem[{{Patrick} {et~al.}(2015){Patrick}, {Evans}, {Davies}, {Kudritzki},
  {Gazak}, {Bergemann}, {Plez}, \& {Ferguson}}]{patrick2015}
{Patrick}, L.~R., {Evans}, C.~J., {Davies}, B., {et~al.} 2015, \apj, 803, 14,
  \dodoi{10.1088/0004-637X/803/1/14}

\bibitem[{{Ramstedt} {et~al.}(2006){Ramstedt}, {Sch{\"o}ier}, {Olofsson}, \&
  {Lundgren}}]{ramstedt2006}
{Ramstedt}, S., {Sch{\"o}ier}, F.~L., {Olofsson}, H., \& {Lundgren}, A.~A.
  2006, \aap, 454, L103, \dodoi{10.1051/0004-6361:20065285}

\bibitem[{{Reimers}(1975)}]{reimers1975circumstellar}
{Reimers}, D. 1975, Memoires of the Societe Royale des Sciences de Liege, 8,
  369

\bibitem[{Ren {et~al.}(2021)Ren, Jiang, Yang, Wang, Jian, \& Ren}]{ren2020}
Ren, Y., Jiang, B., Yang, M., {et~al.} 2021, The Astrophysical Journal, 907,
  18, \dodoi{10.3847/1538-4357/abcda5}

\bibitem[{{Riebel} {et~al.}(2012){Riebel}, {Srinivasan}, {Sargent}, \&
  {Meixner}}]{Riebel2012}
{Riebel}, D., {Srinivasan}, S., {Sargent}, B., \& {Meixner}, M. 2012, \apj,
  753, 71, \dodoi{10.1088/0004-637X/753/1/71}

\bibitem[{{Rieke} {et~al.}(2004){Rieke}, {Young}, {Engelbracht}, {Kelly},
  {Low}, {Haller}, {Beeman}, {Gordon}, {Stansberry}, {Misselt}, {Cadien},
  {Morrison}, {Rivlis}, {Latter}, {Noriega-Crespo}, {Padgett}, {Stapelfeldt},
  {Hines}, {Egami}, {Muzerolle}, {Alonso-Herrero}, {Blaylock}, {Dole}, {Hinz},
  {Le Floc'h}, {Papovich}, {P{\'e}rez-Gonz{\'a}lez}, {Smith}, {Su}, {Bennett},
  {Frayer}, {Henderson}, {Lu}, {Masci}, {Pesenson}, {Rebull}, {Rho}, {Keene},
  {Stolovy}, {Wachter}, {Wheaton}, {Werner}, \& {Richards}}]{mips}
{Rieke}, G.~H., {Young}, E.~T., {Engelbracht}, C.~W., {et~al.} 2004, \apjs,
  154, 25, \dodoi{10.1086/422717}

\bibitem[{{Rouleau} \& {Martin}(1991)}]{rouleau1991shape}
{Rouleau}, F., \& {Martin}, P.~G. 1991, \apj, 377, 526, \dodoi{10.1086/170382}

\bibitem[{{Sargent} {et~al.}(2010){Sargent}, {Srinivasan}, {Meixner}, {Kemper},
  {Tielens}, {Speck}, {Matsuura}, {Bernard}, {Hony}, {Gordon}, {Indebetouw},
  {Marengo}, {Sloan}, \& {Woods}}]{sargent2010}
{Sargent}, B.~A., {Srinivasan}, S., {Meixner}, M., {et~al.} 2010, \apj, 716,
  878, \dodoi{10.1088/0004-637X/716/1/878}

\bibitem[{{Schlegel} {et~al.}(1998){Schlegel}, {Finkbeiner}, \&
  {Davis}}]{sfd98}
{Schlegel}, D.~J., {Finkbeiner}, D.~P., \& {Davis}, M. 1998, \apj, 500, 525,
  \dodoi{10.1086/305772}

\bibitem[{{Scicluna} {et~al.}(2015){Scicluna}, {Siebenmorgen}, {Wesson},
  {Blommaert}, {Kasper}, {Voshchinnikov}, \& {Wolf}}]{scicluna2015}
{Scicluna}, P., {Siebenmorgen}, R., {Wesson}, R., {et~al.} 2015, \aap, 584,
  L10, \dodoi{10.1051/0004-6361/201527563}

\bibitem[{{Sibbons} {et~al.}(2012){Sibbons}, {Ryan}, {Cioni}, {Irwin}, \&
  {Napiwotzki}}]{sibbons2012}
{Sibbons}, L.~F., {Ryan}, S.~G., {Cioni}, M. R.~L., {Irwin}, M., \&
  {Napiwotzki}, R. 2012, \aap, 540, A135, \dodoi{10.1051/0004-6361/201118365}

\bibitem[{{Smartt}(2009)}]{smartt2009}
{Smartt}, S.~J. 2009, \araa, 47, 63,
  \dodoi{10.1146/annurev-astro-082708-101737}

\bibitem[{{Srinivasan} {et~al.}(2011){Srinivasan}, {Sargent}, \&
  {Meixner}}]{grams}
{Srinivasan}, S., {Sargent}, B.~A., \& {Meixner}, M. 2011, \aap, 532, A54,
  \dodoi{10.1051/0004-6361/201117033}

\bibitem[{{Sylvester} {et~al.}(1998){Sylvester}, {Skinner}, \&
  {Barlow}}]{sylvester1998}
{Sylvester}, R.~J., {Skinner}, C.~J., \& {Barlow}, M.~J. 1998, \mnras, 301,
  1083, \dodoi{10.1046/j.1365-8711.1998.02078.x}

\bibitem[{{Taylor}(2005)}]{topcat}
{Taylor}, M.~B. 2005, in Astronomical Society of the Pacific Conference Series,
  Vol. 347, Astronomical Data Analysis Software and Systems XIV, ed.
  P.~{Shopbell}, M.~{Britton}, \& R.~{Ebert}, 29

\bibitem[{{U} {et~al.}(2009){U}, {Urbaneja}, {Kudritzki}, {Jacobs}, {Bresolin},
  \& {Przybilla}}]{u2009}
{U}, V., {Urbaneja}, M.~A., {Kudritzki}, R.-P., {et~al.} 2009, \apj, 704, 1120,
  \dodoi{10.1088/0004-637X/704/2/1120}

\bibitem[{{Ueta} \& {Meixner}(2003)}]{ueta2003}
{Ueta}, T., \& {Meixner}, M. 2003, \apj, 586, 1338, \dodoi{10.1086/367818}

\bibitem[{{van Loon} {et~al.}(2005){van Loon}, {Cioni}, {Zijlstra}, \&
  {Loup}}]{van2005empirical}
{van Loon}, J.~T., {Cioni}, M. R.~L., {Zijlstra}, A.~A., \& {Loup}, C. 2005,
  \aap, 438, 273, \dodoi{10.1051/0004-6361:20042555}

\bibitem[{{Verhoelst} {et~al.}(2009){Verhoelst}, {van der Zypen}, {Hony},
  {Decin}, {Cami}, \& {Eriksson}}]{verhoelst2009}
{Verhoelst}, T., {van der Zypen}, N., {Hony}, S., {et~al.} 2009, \aap, 498,
  127, \dodoi{10.1051/0004-6361/20079063}

\bibitem[{{Wang} \& {Chen}(2019)}]{wang2019}
{Wang}, S., \& {Chen}, X. 2019, \apj, 877, 116,
  \dodoi{10.3847/1538-4357/ab1c61}

\bibitem[{{Werner} {et~al.}(2004){Werner}, {Roellig}, {Low}, {Rieke}, {Rieke},
  {Hoffmann}, {Young}, {Houck}, {Brandl}, {Fazio}, {Hora}, {Gehrz}, {Helou},
  {Soifer}, {Stauffer}, {Keene}, {Eisenhardt}, {Gallagher}, {Gautier}, {Irace},
  {Lawrence}, {Simmons}, {Van Cleve}, {Jura}, {Wright}, \&
  {Cruikshank}}]{spitzer}
{Werner}, M.~W., {Roellig}, T.~L., {Low}, F.~J., {et~al.} 2004, \apjs, 154, 1,
  \dodoi{10.1086/422992}

\bibitem[{{Wright} {et~al.}(2010){Wright}, {Eisenhardt}, {Mainzer}, {Ressler},
  {Cutri}, {Jarrett}, {Kirkpatrick}, {Padgett}, {McMillan}, {Skrutskie},
  {Stanford}, {Cohen}, {Walker}, {Mather}, {Leisawitz}, {Gautier}, {McLean},
  {Benford}, {Lonsdale}, {Blain}, {Mendez}, {Irace}, {Duval}, {Liu}, {Royer},
  {Heinrichsen}, {Howard}, {Shannon}, {Kendall}, {Walsh}, {Larsen}, {Cardon},
  {Schick}, {Schwalm}, {Abid}, {Fabinsky}, {Naes}, \& {Tsai}}]{wise}
{Wright}, E.~L., {Eisenhardt}, P. R.~M., {Mainzer}, A.~K., {et~al.} 2010, \aj,
  140, 1868, \dodoi{10.1088/0004-6256/140/6/1868}

\bibitem[{{Xue} {et~al.}(2016){Xue}, {Jiang}, {Gao}, {Liu}, {Wang}, \&
  {Li}}]{xue2016}
{Xue}, M., {Jiang}, B.~W., {Gao}, J., {et~al.} 2016, \apjs, 224, 23,
  \dodoi{10.3847/0067-0049/224/2/23}

\bibitem[{{Yang} {et~al.}(2018){Yang}, {Bonanos}, {Jiang}, {Gao}, {Xue},
  {Wang}, {Lam}, {Spetsieri}, {Ren}, \& {Gavras}}]{ym2018}
{Yang}, M., {Bonanos}, A.~Z., {Jiang}, B.-W., {et~al.} 2018, \aap, 616, A175,
  \dodoi{10.1051/0004-6361/201832833}

\bibitem[{{Yang} {et~al.}(2019){Yang}, {Bonanos}, {Jiang}, {Gao}, {Gavras},
  {Maravelias}, {Ren}, {Wang}, {Xue}, {Tramper}, {Spetsieri}, \&
  {Pouliasis}}]{ym2019}
---. 2019, \aap, 629, A91, \dodoi{10.1051/0004-6361/201935916}

\bibitem[{{Yang} {et~al.}(2020){Yang}, {Bonanos}, {Jiang}, {Gao}, {Gavras},
  {Maravelias}, {Wang}, {Chen}, {Tramper}, {Ren}, {Spetsieri}, \&
  {Xue}}]{ym2020}
---. 2020, \aap, 639, A116, \dodoi{10.1051/0004-6361/201937168}

\end{thebibliography}
\bibliographystyle{aasjournal}

\begin{figure}[ht!]
    \centering
    \includegraphics[width=7in]{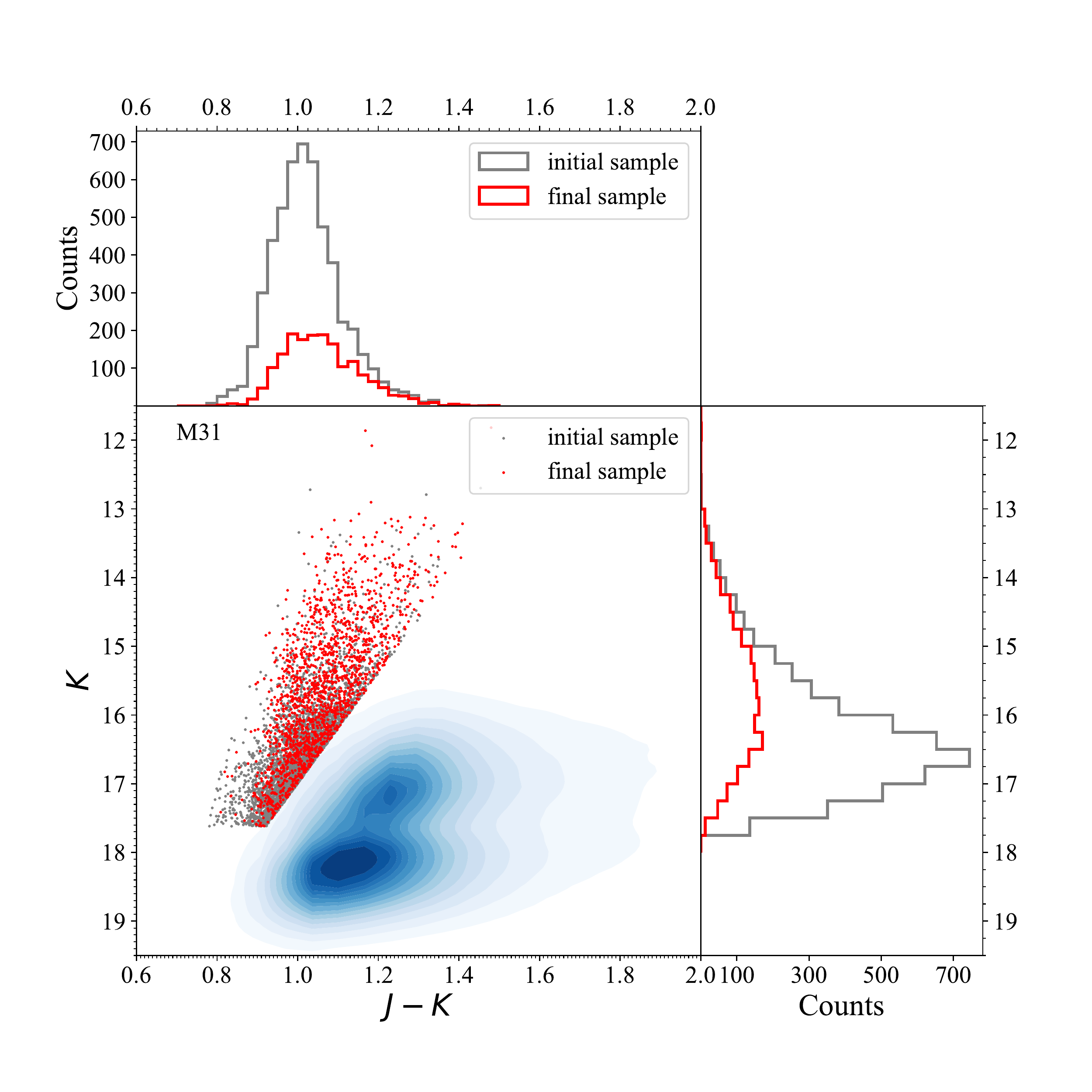}
    \caption{The initial (grey dots) and final (red dots) sample of RSGs in the $J-K/K$ diagram for M31 as well as the histogram of $J-K$ and $K$. Also plotted is the density contour of RGB and AGB stars for reference.   \label{jkm31}}
\end{figure}

\begin{figure}[ht!]
    \centering
    \includegraphics[width=7in]{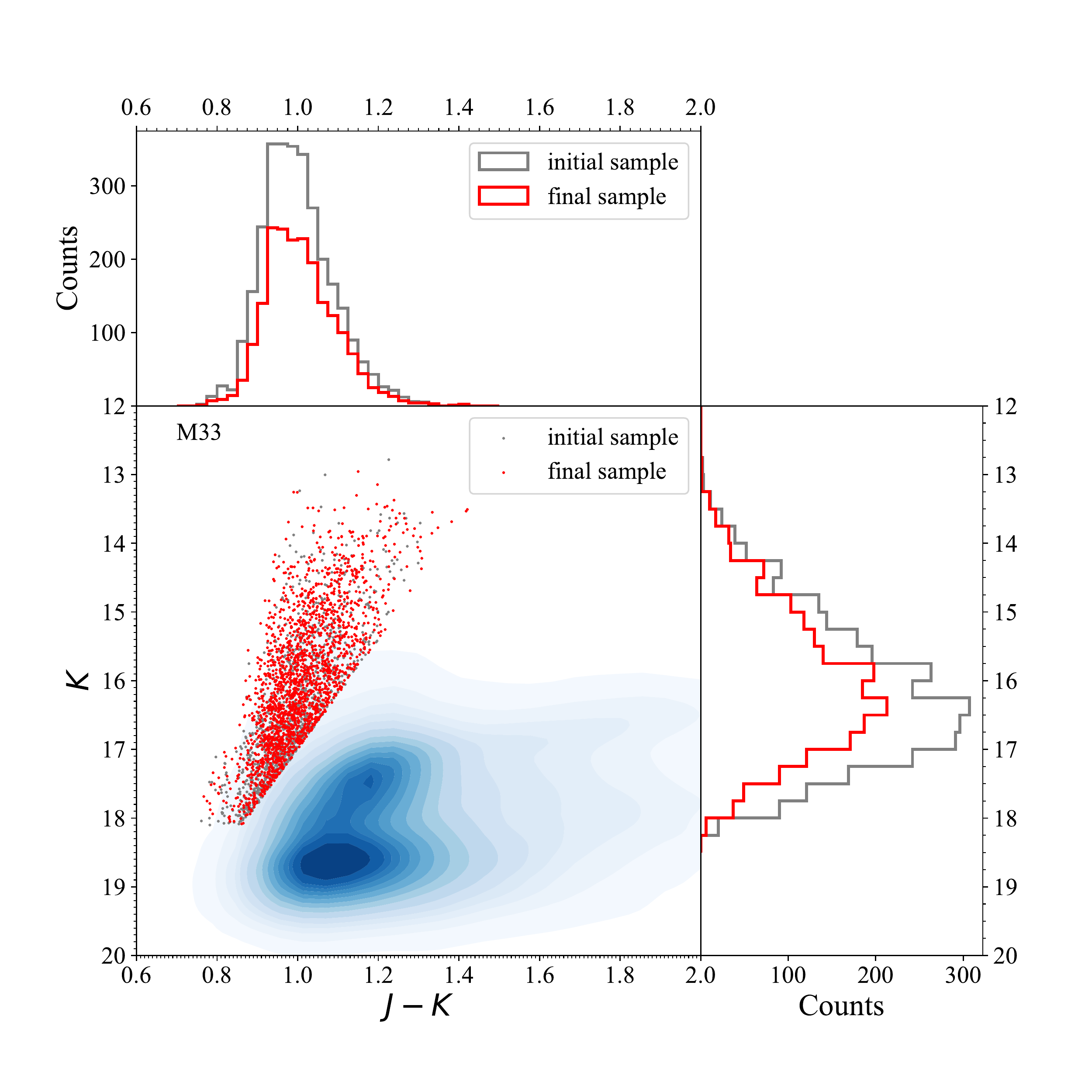}
    \caption{The RSGs  in the $J-K/K$ diagram for M33, the symbols are the same as in Figure \ref{jkm31}.  \label{jkm33}}
\end{figure}
\begin{figure}[ht!]
    \centering
    \includegraphics[width=7in]{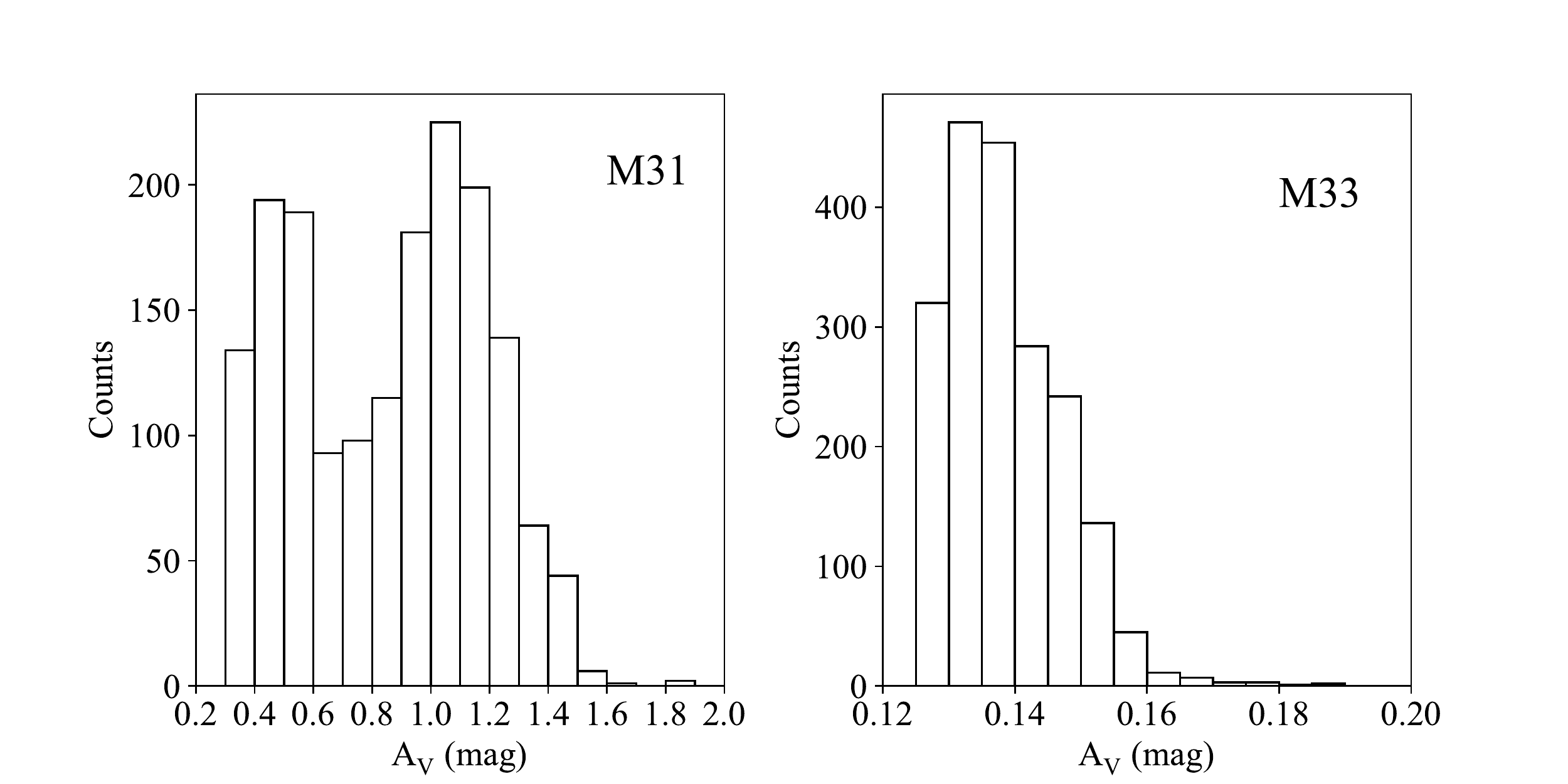}
    \caption{The $A_V$ distribution of RSGs in M31 and M33.   \label{av}}
\end{figure}

\begin{figure}[ht!]
    \centering
    \includegraphics[width=7in]{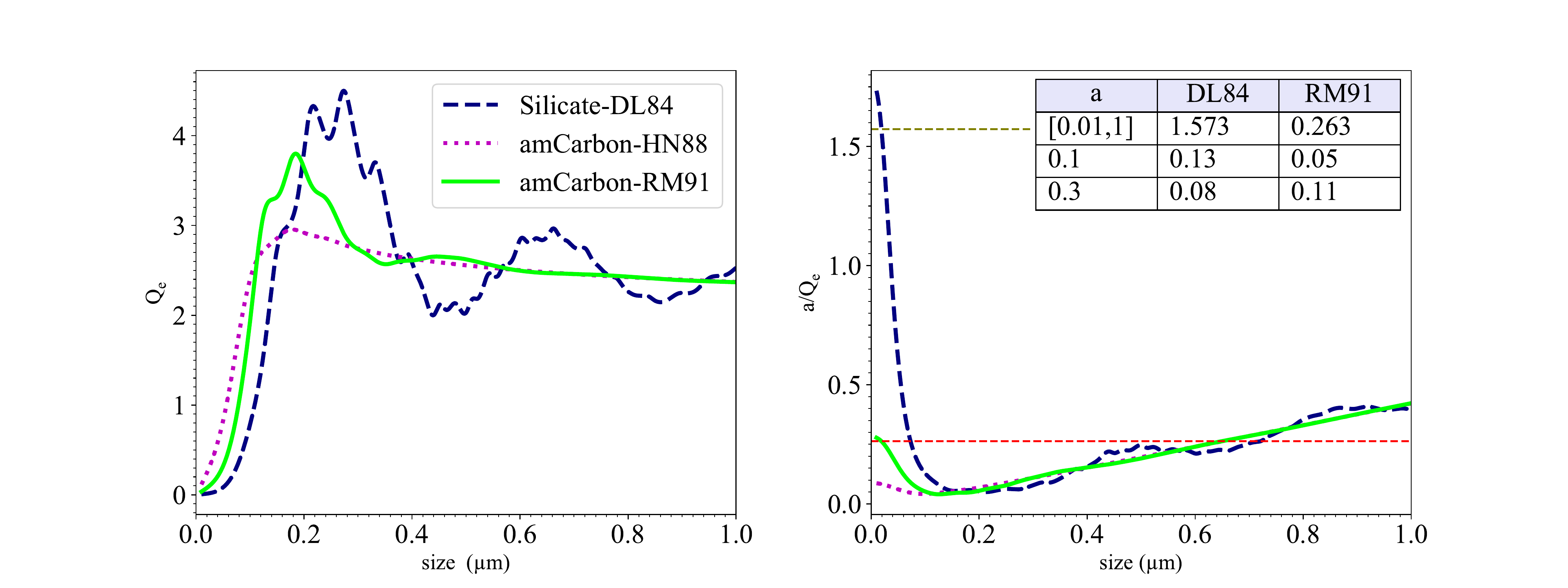}
    \caption{The change of the extinction coefficient $Q_e$ (left) and the ratio $a/Q_e$ (right) in the $V$ band  with the radius of the dust grain $a$ for three dust species, i.e. amorphous silicate from DL84, amorphous carbon from RM91 and HN88 respectively. The inset shows the value of $a/Q_e$ for an MRN  size distribution adopted in this work, and for a unique radius of $a=0.1\mu$m and $a=0.3\mu$m used in previous works.  \label{qe}}
\end{figure}

\begin{figure}[ht!]
    \centering
    \includegraphics[width=7in]{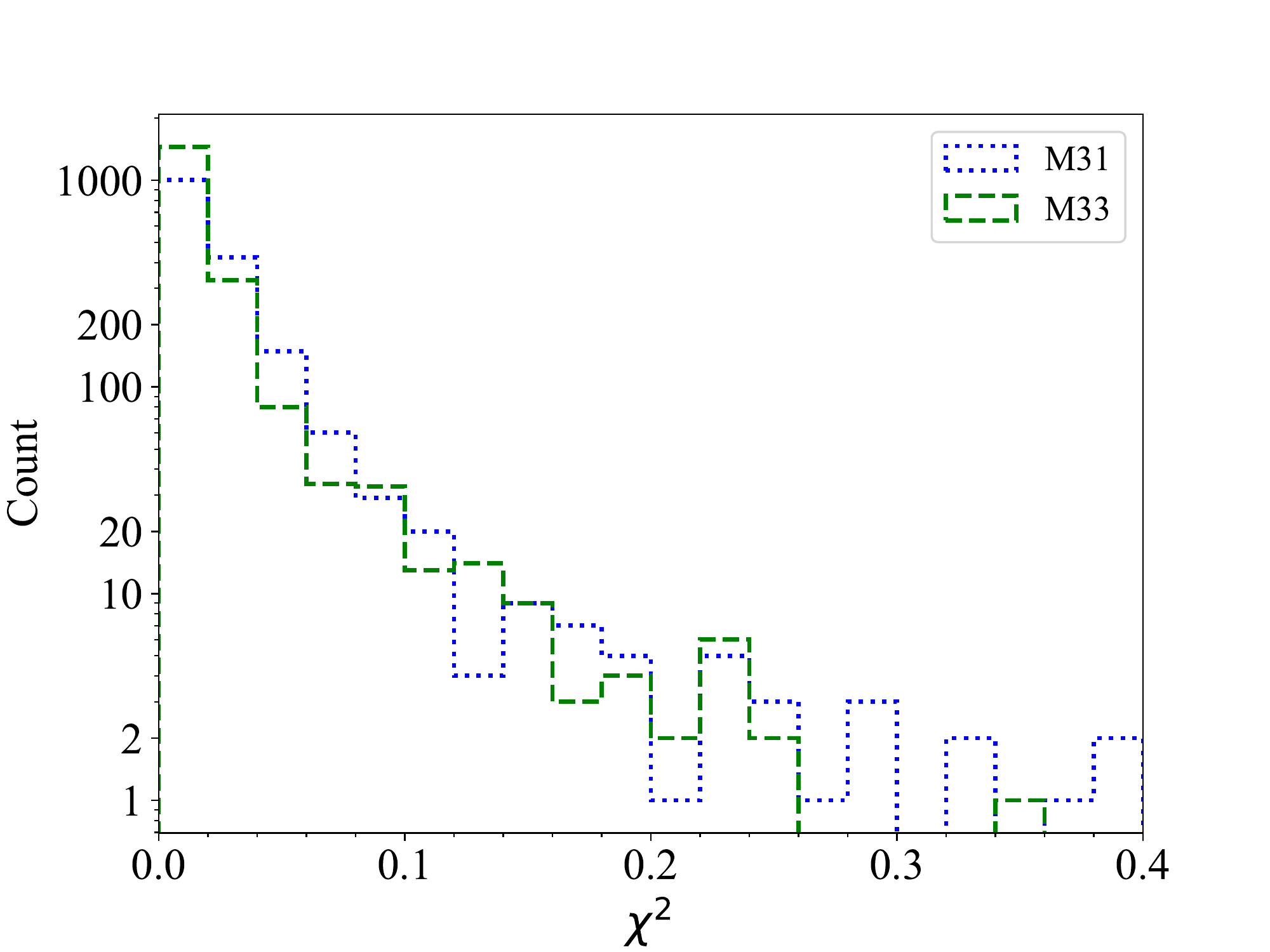}
    \caption{Distribution of $\chi^2$ of the model fitting for M31 (blue line) and  M33 (green line). \label{chi2}}
\end{figure}

\newpage

\begin{figure}[ht]
    \centering
    \includegraphics[width=7in]{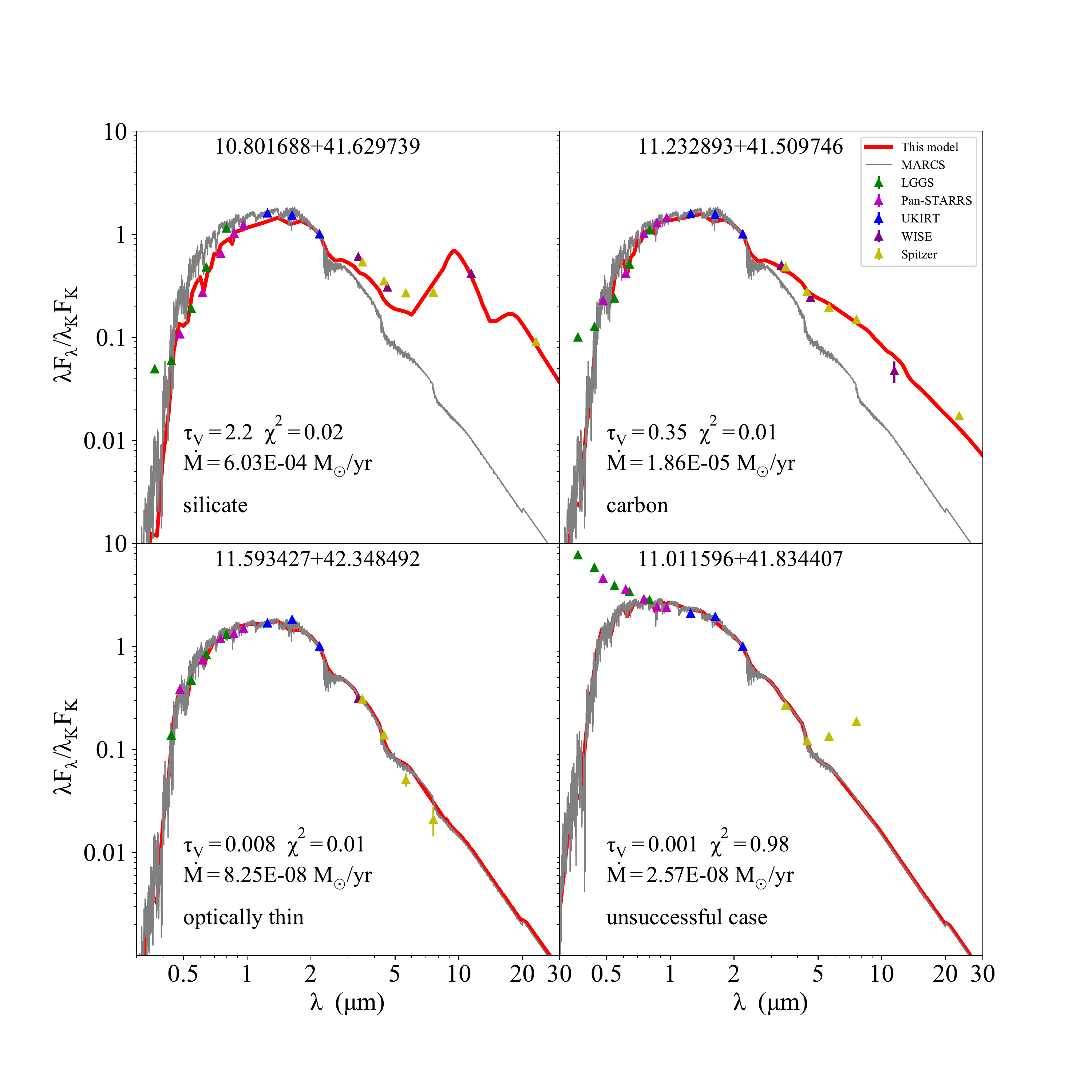}
    \caption{The examples of four typical dust model fittings (red line) to the spectral energy distributions, in a clockwise direction are the cases of amorphous silicate, amorphous carbon, optically thin and unsuccessful respectively. Also shown are the MARCS photospheric model spectrum (grey line), the photometric results from various surveys (triangles), the key model fitting parameters as well as the coordinate name of the source. \label{spec}}
\end{figure}

\begin{figure}[ht!]
    \centering
    \includegraphics[width=7in]{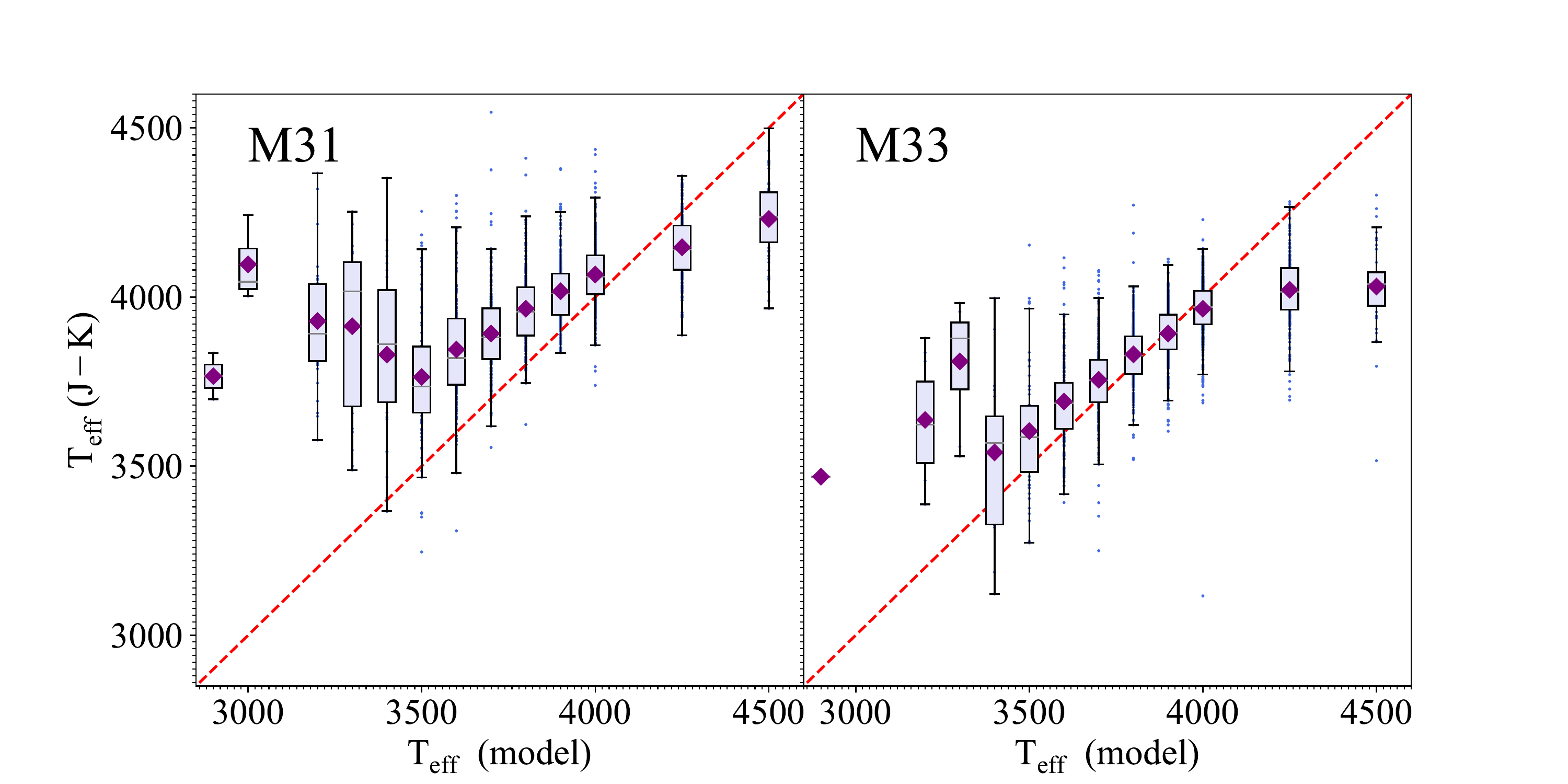}
    \caption{Comparison of the effective temperature derived from the model fitting and from the color index $J - K$. \label{teff-teff}}
\end{figure}

\begin{figure}[ht!]
    \centering
    \includegraphics[width=7in]{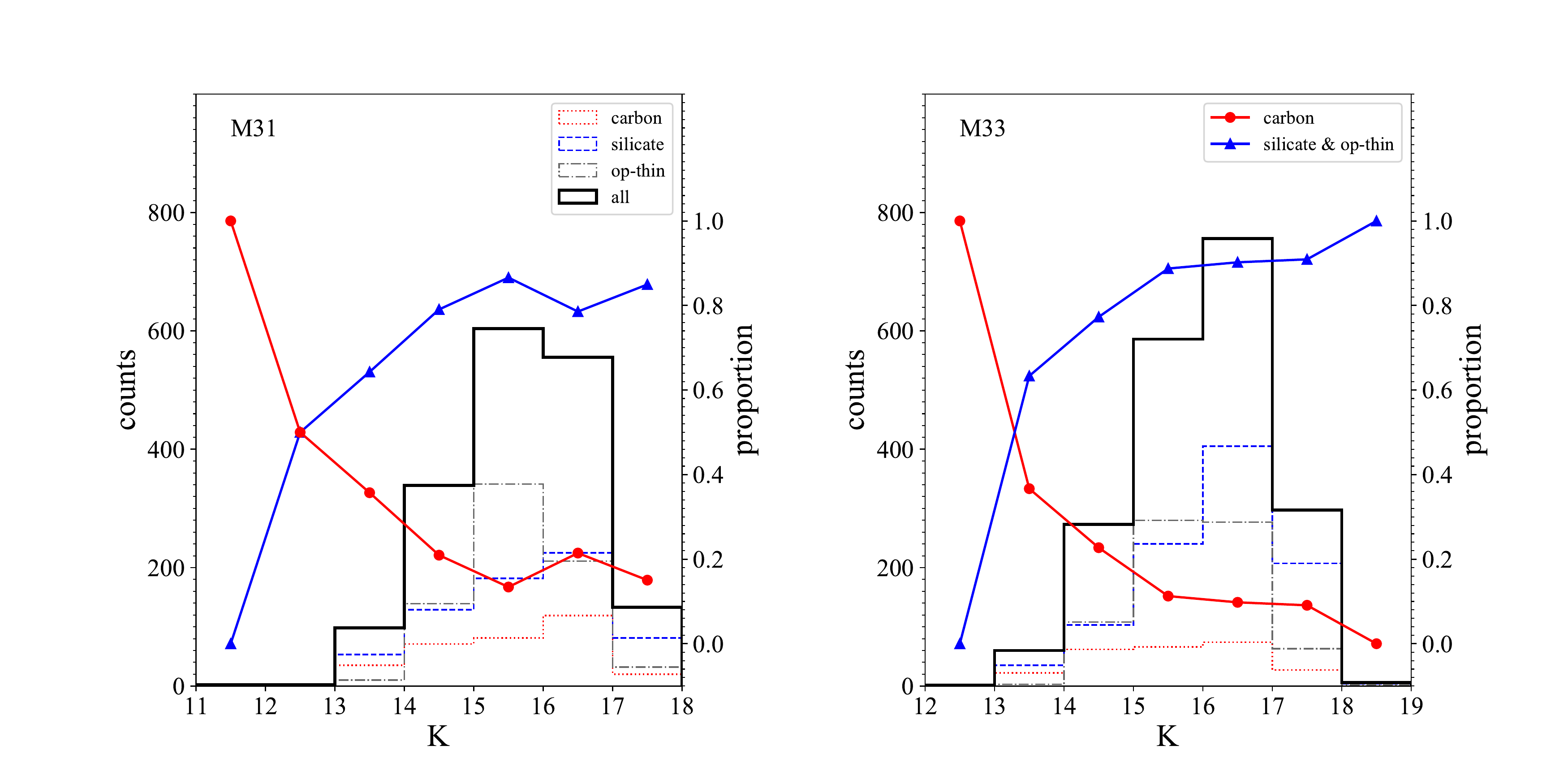}
    \caption{The change of carbon RSGs proportion with the $K$ magnitude, together with the histogram of RSGs with silicate and carbon dust respectively.  \label{kcp}}
\end{figure}

\begin{figure}[ht!]
    \centering
    \includegraphics[width=7in]{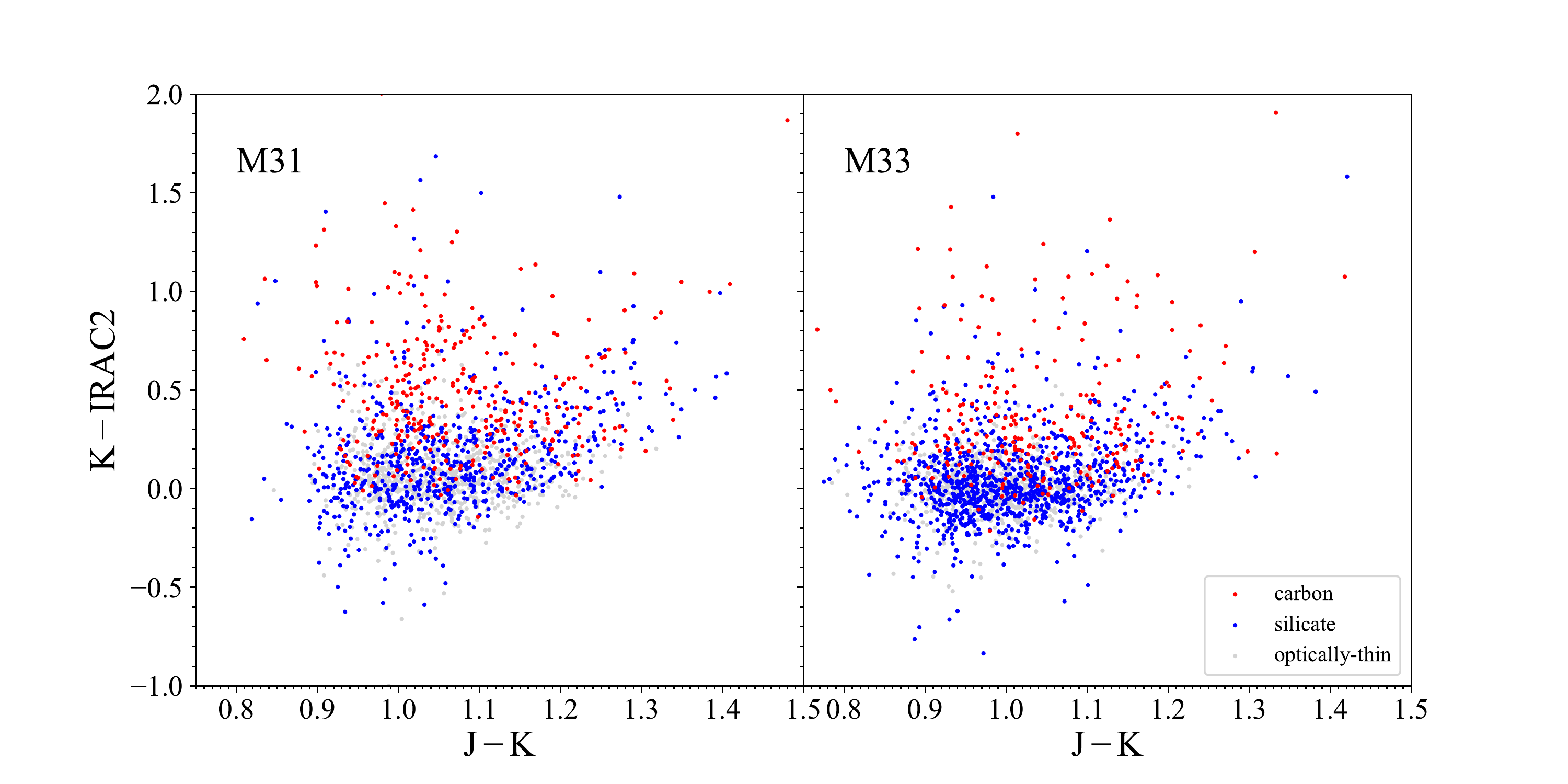}
    \caption{The $J - K / K - IRAC2$ diagram for RSGs with amorphous silicate, amorphous carbon dust as well as the optically thin sources.  \label{JK-K4}}
\end{figure}

\begin{figure}[ht!]
    \centering
    \includegraphics[width=7in]{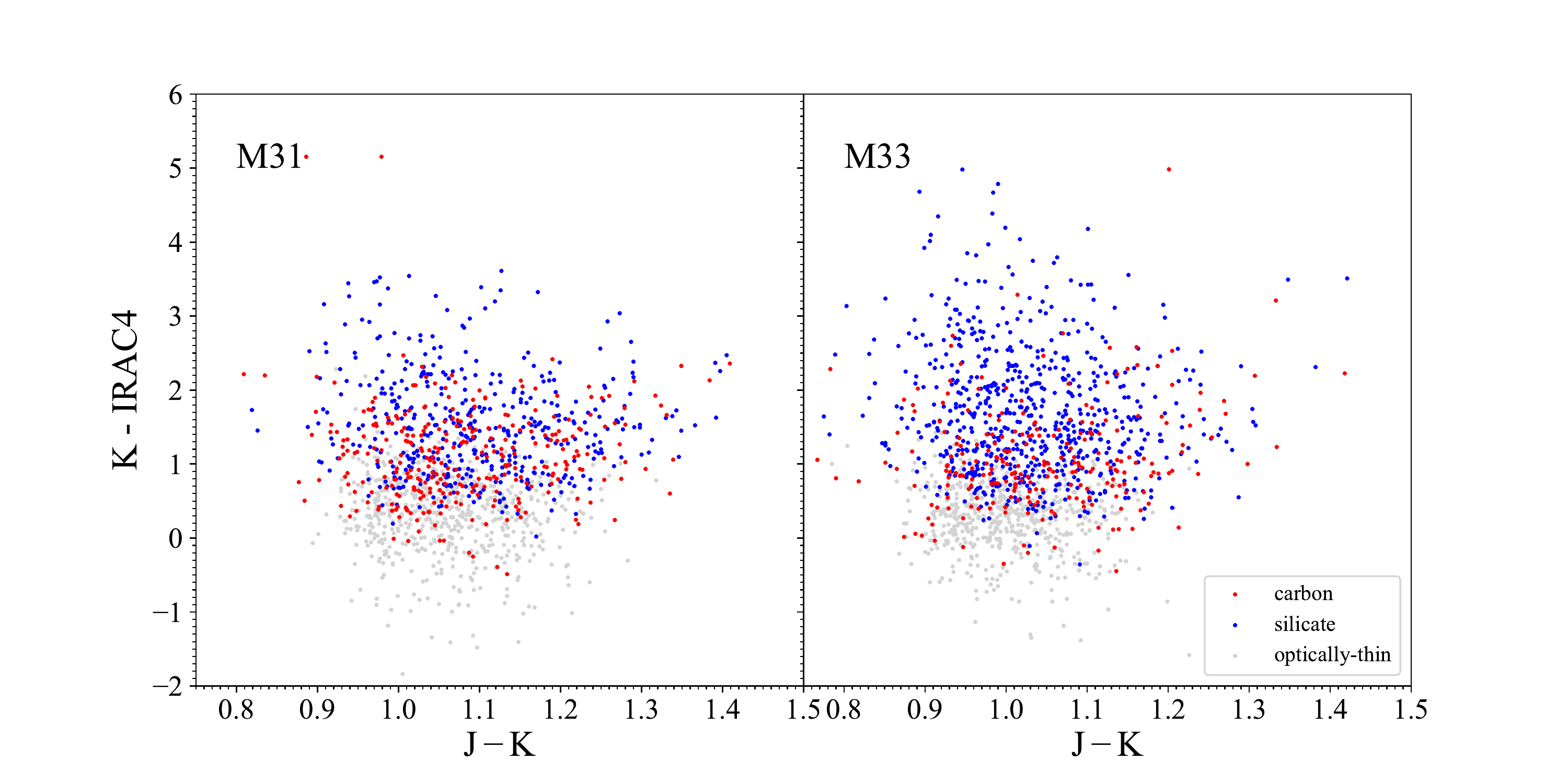}
    \caption{The $J - K / K - IRAC4$ diagram for RSGs with amorphous silicate, amorphous carbon dust as well as the optically thin sources.  \label{JK-K8}}
\end{figure}

\begin{figure}[ht!]
    \centering
    \includegraphics[width=7in]{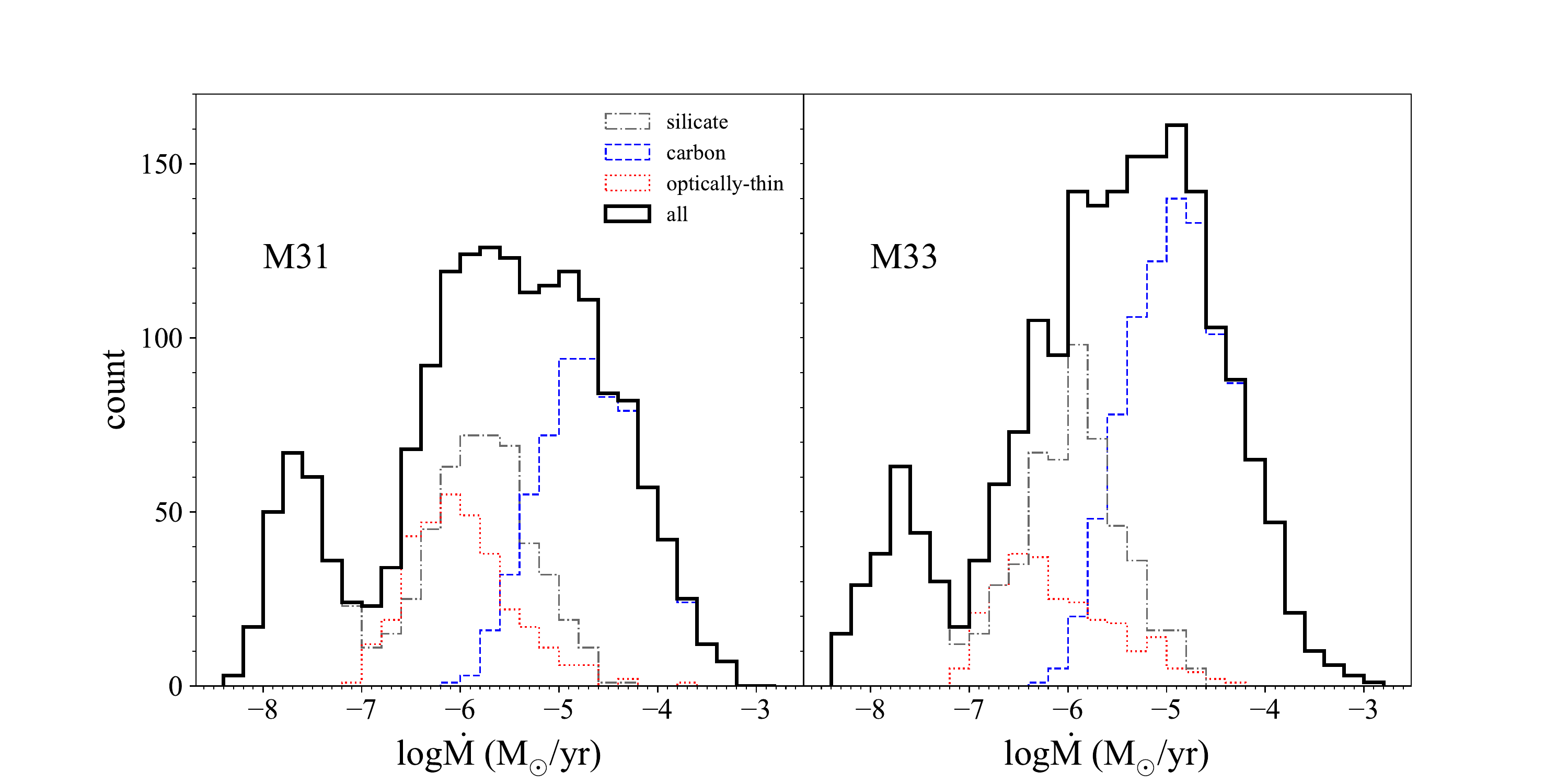}
    \caption{The distribution of mass loss rates for RSGs in M31 (left) and M33 (right). \label{histmlr}}
\end{figure}

\begin{figure}[ht!]
    \centering
    \includegraphics[width=7in]{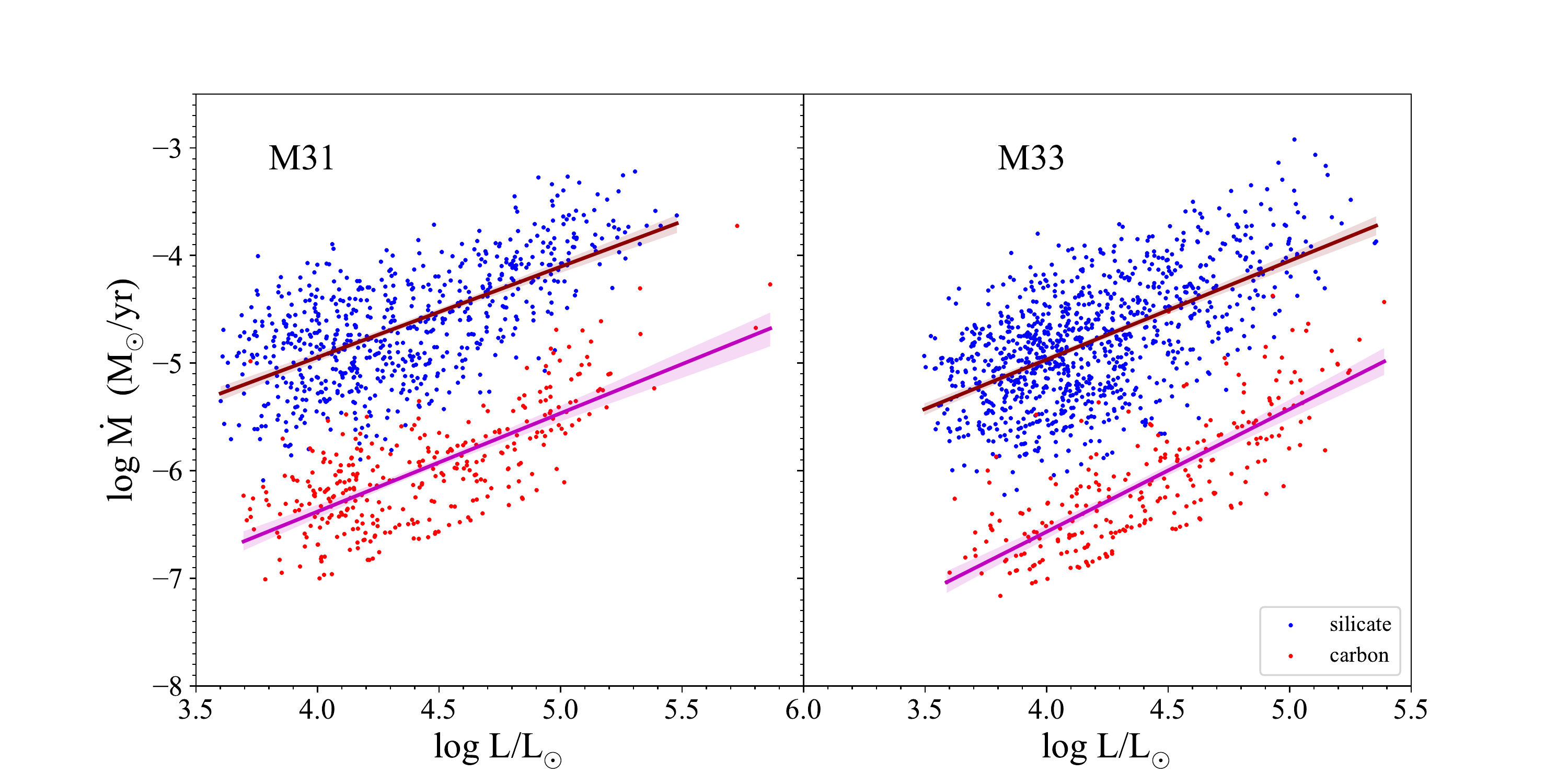}
    \caption{The relation of mass loss rate to the luminosity. \label{mlr-lum}}
\end{figure}

\begin{figure}[ht!]
    \includegraphics[width=7in]{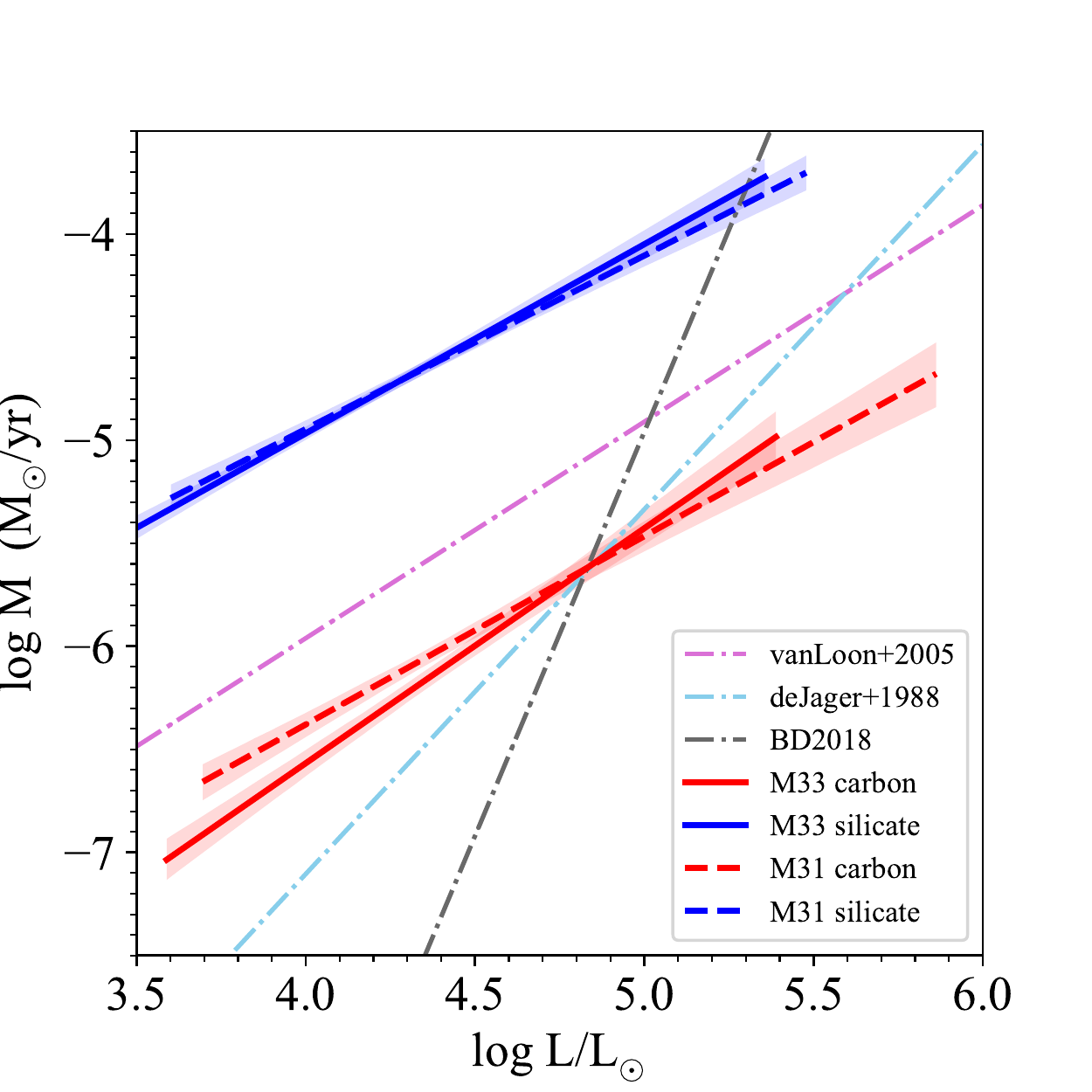}
    \caption{Comparison of the relation of mass loss rate to the luminosity with the works of \citet{dj1988}, \citet{van2005empirical} and \citet{beasor2018evolution}. \label{mlr-lum-comp}}
\end{figure}

\begin{figure}[ht!]
    \centering
    \includegraphics[width=7in]{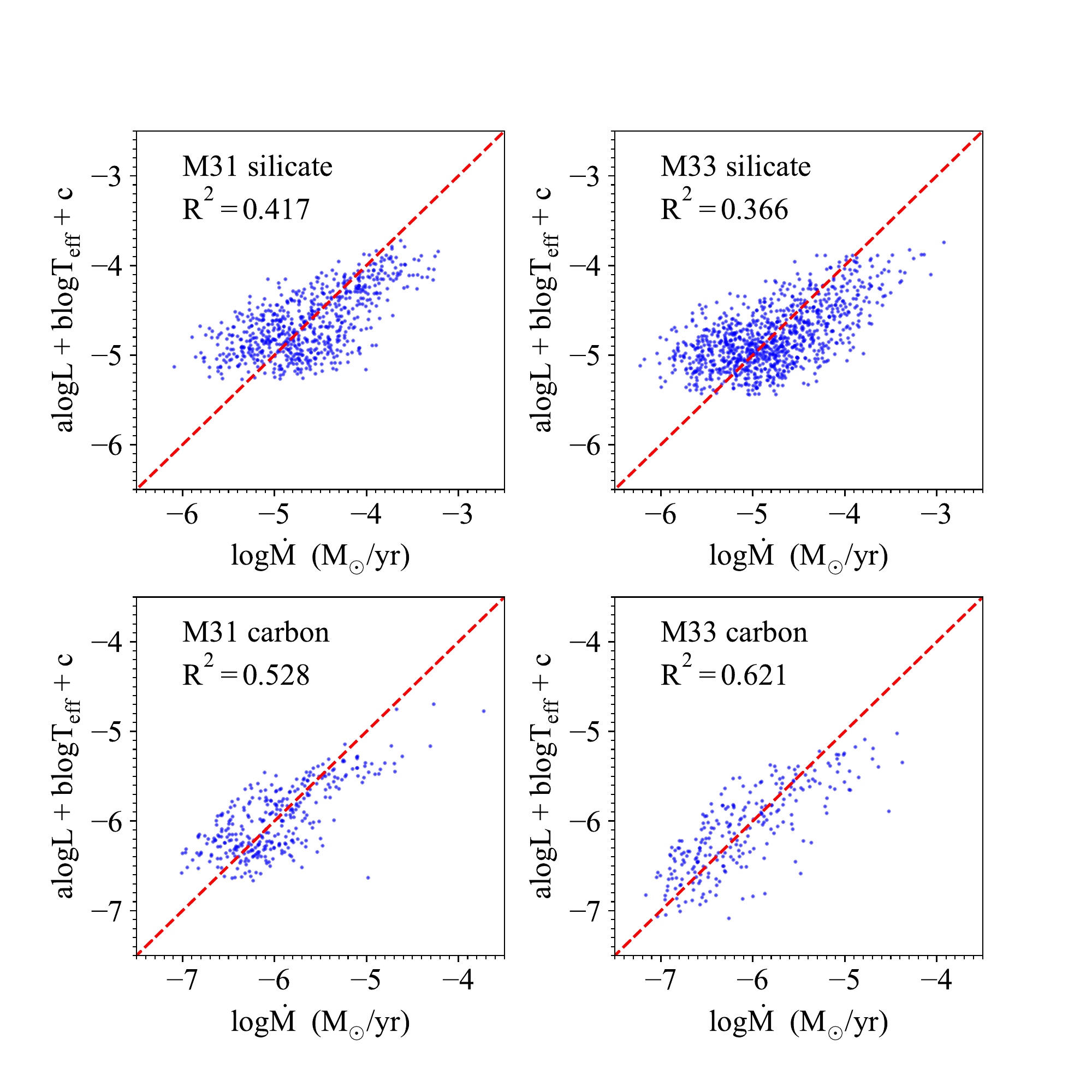}
    \caption{The relation of mass loss rate with the luminosity and the effective temperature, and the red line marks  $Y=X$. \label{mlr-teff}}
\end{figure}

\begin{figure}
    \centering
    \includegraphics[width=7in]{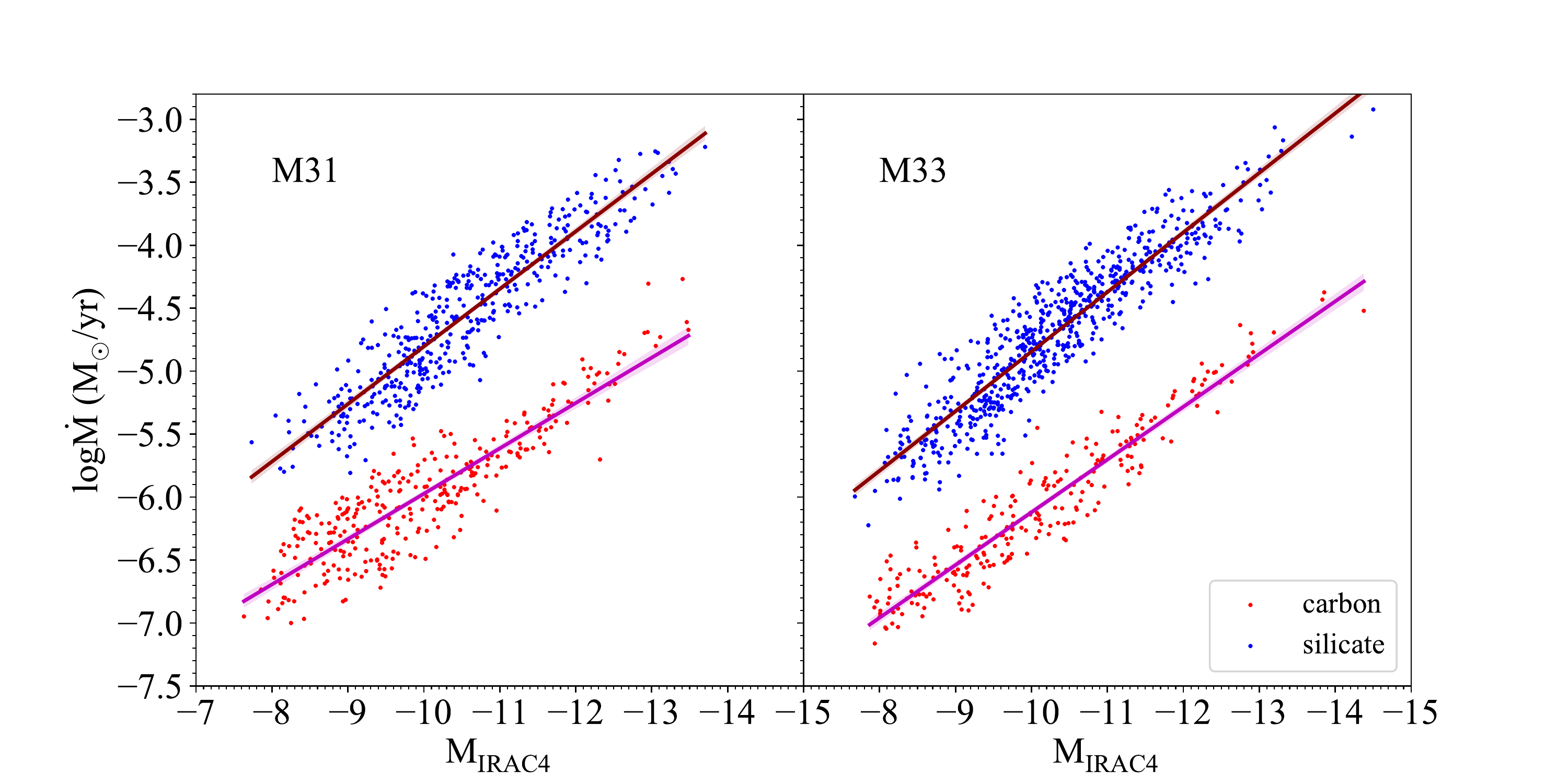}
    \caption{The relation of mass loss rate to the \emph{Spitzer}/IRAC4 absolute magnitude.}
    \label{mlr_I4}
\end{figure}

\begin{figure}
    \centering
    \includegraphics[width=7in]{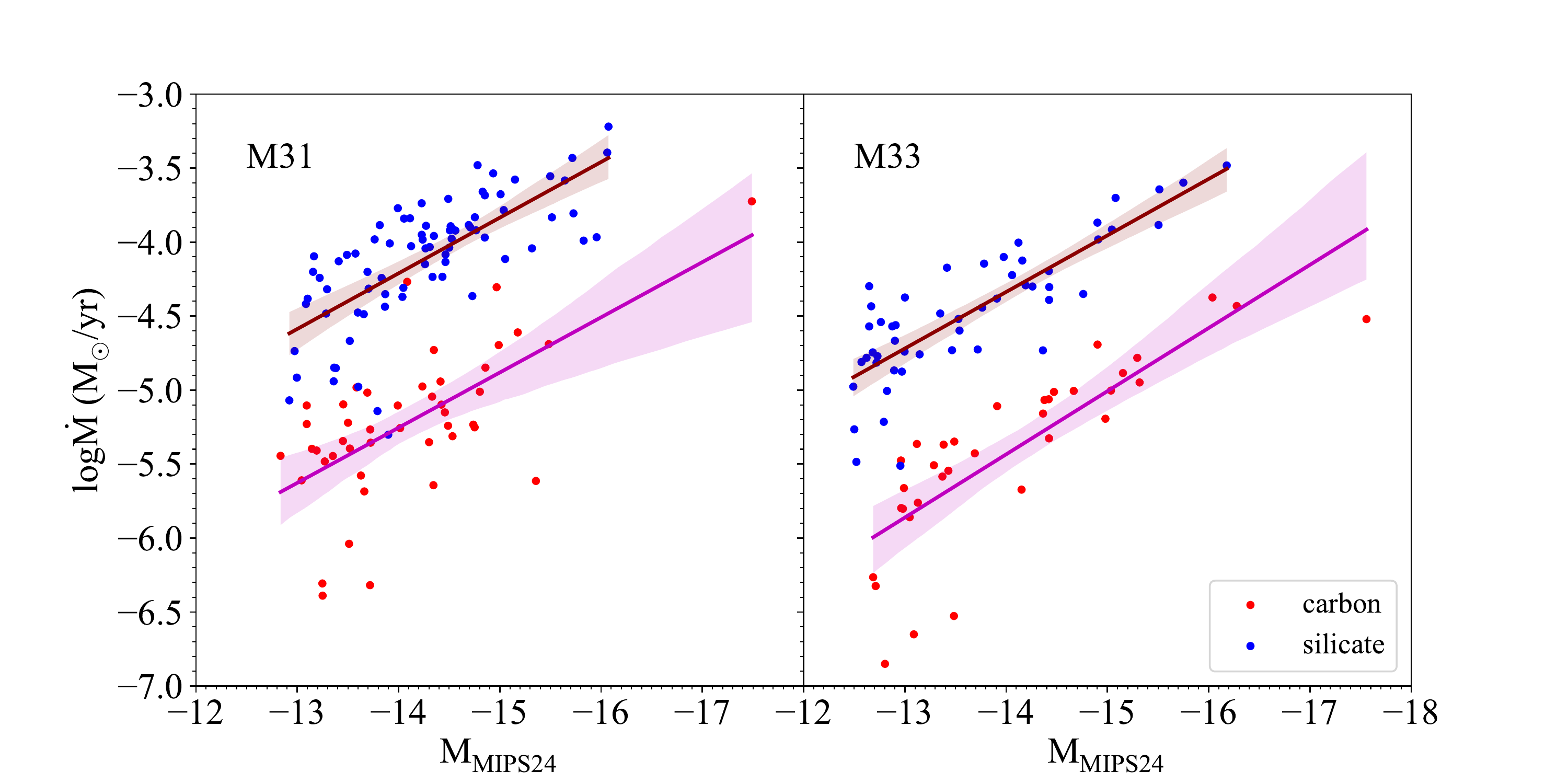}
    \caption{The relation of mass loss rate to the \emph{Spitzer}/MIPS24 absolute magnitude. }
    \label{mlr_M1}
\end{figure}

\begin{figure}[ht!]
    \includegraphics[width=7in]{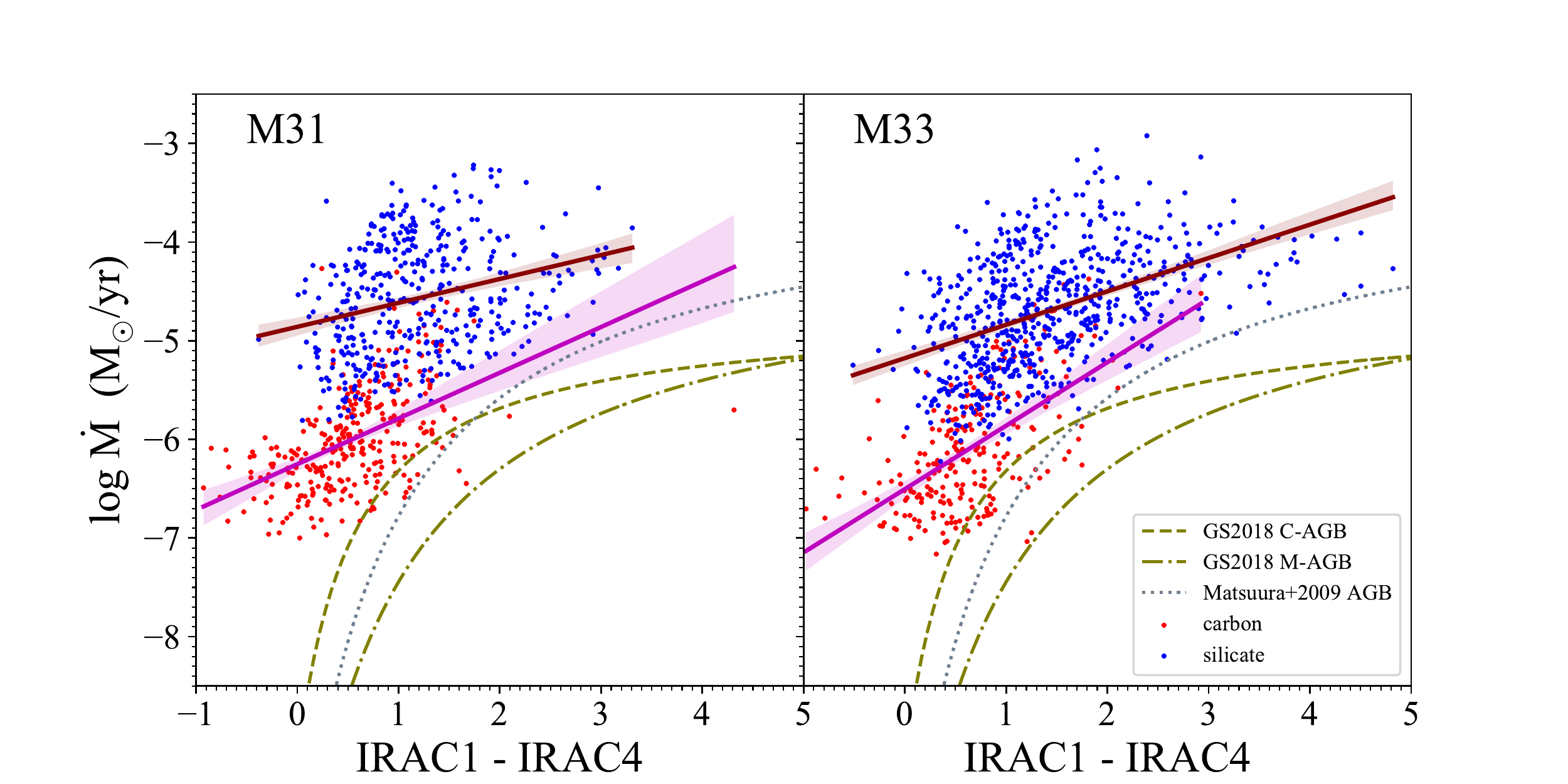}
    \caption{The relation of mass loss rate to the color index $IRAC1-IRAC4$, which is compared with that from \citet{Groenewegen2018} and \citet{matsuura2009}. \label{mlr-I1-I4}}
\end{figure}

\iffalse
\fi
\clearpage

\begin{deluxetable*}{cccc}
\tablenum{1}
\tablecaption{Number of RSGs in M31 and M33 \label{count} }
\tablewidth{20pt}
\tablehead{\colhead{} & \colhead{M31+M33} & \colhead{M31}& \colhead{M33}}
\startdata
{  Paper I}    & {  8254}    & {  5253}    & {  3001}    \\ \hline
{  sample}     & {  3712}    & {  1733}    & {  1979}    \\ \hline
{  carbon}     & {  581}     & {  329}     & {  252}      \\ \hline
{  silicate}   & {  1665}     & {  671}     & {  994}    \\ \hline
{  op-thin}    & {  1466}    & {  733}     & {  733} \\ \hline
{  Carbon\%}        & { 16\%} & {  19\%}   & {  13\%}  \\ \hline
{  Op-thin\%} & {  39\%} & {  42\%} & {  37\%} \\
\enddata
\end{deluxetable*}

\begin{deluxetable*}{c|ccc|ccc}
\tablenum{2}
\tablecaption{Mass loss rates of RSGs in M31 and M33  (M$_{\odot}$/yr) }\label{mean}
\tablewidth{20pt}
\tablehead{
\colhead{} & \multicolumn3c{M31} & \multicolumn3c{M33}  \\ \hline
\colhead{} & \colhead{carbon} & \colhead {silicate} &\colhead{all} & \colhead{carbon} & \colhead {silicate} &\colhead{all}}
\startdata
mean & 2.92E-06 & 4.86E-05 & 2.02E-05 & 2.34E-06 & 3.82E-05 & 2.01E-05 \\ \hline
median & 8.92E-07 & 2.11E-05 & 2.66E-06 & 6.13E-07 & 1.49E-05 & 3.45E-06 \\ \hline
min & 9.80E-08 & 8.14E-07 & 4.29E-09 & 6.87E-08 & 5.97E-07 & 3.54E-09 \\ \hline
max & 1.88E-04 & 6.03E-04 & 6.03E-04 & 4.22E-05 & 1.20E-03 & 1.20E-03 \\ \hline
sum & 9.59E-04 & 3.26E-02 & 3.51E-02 & 5.91E-04 & 3.80E-02 & 3.97E-02 \\
\enddata

\end{deluxetable*}

\begin{deluxetable*}{c|cc|cc}
\tablenum{3}
\tablecaption{The Spearman's rank correlation coefficient  of the logarithmic  mass loss rate with stellar luminosity, infrared magnitude and colors \label{spearman} }
\tablewidth{20pt}
\tablehead{
\colhead{} & \multicolumn2c{M31} & \multicolumn2c{M33}  \\
\colhead{} & \colhead{silicate} & \colhead {carbon} &\colhead{silicate} & \colhead {carbon}}
\startdata
    log$L$  & 0.59 & 0.66 & 0.54 & 0.78  \\  \hline
    $M_{\rm IRAC4}$ & -0.92 & -0.86 & -0.93 & -0.94   \\ \hline
    $M_{\rm MIPS24}$    & -0.76 & -0.56 & -0.80 & -0.90     \\ \hline
    IRAC1-IRAC4 & 0.29&  0.57& 0.50&  0.56\\
\enddata
\end{deluxetable*}

\begin{deluxetable*}{ccccccccccccc}
\tablenum{4}
\tablecaption{Catalog of the stellar and dust parameters for red supergiants in M31}\label{cat1}
\tablewidth{0pt}
\tablehead{
\colhead{index} & \colhead{RA} & \colhead{DEC} & \colhead{galaxy} & \colhead{KMag} & \colhead{$\rm KMag\_{Err}$} & \colhead{log$L/L_{\odot}$} & \colhead{$T_{\rm eff}$} & \colhead{ $\dot{M}$ } & \colhead{$T_{\rm in}$} & \colhead{$\tau_V$} & \colhead{$\chi^2$} & \colhead{type}\\
\colhead{} & \colhead{(deg)} & \colhead{(deg)} & \colhead{} & \colhead{(mag)} & \colhead{(mag)} & \colhead{} & \colhead{(K)} & \colhead{ ($M_{\odot}$/yr) } & \colhead{(K)} & \colhead{} & \colhead{} & \colhead{}}
\decimalcolnumbers
\startdata
0 & 11.446398 & 42.345503 & M31 & 15.689 & 0.010 & 4.312 & 3880 & 6.90E-07 & 1200 & 0.04 & 0.009 & op-thin     \\ \hline
1 & 11.469417 & 42.345959 & M31 & 14.077 & 0.010 & 4.862 & 3564 & 9.02E-06 & 750  & 0.45 & 0.016 & carbon   \\ \hline
2 & 11.518298 & 42.411805 & M31 & 15.118 & 0.010 & 4.506 & 3762 & 2.66E-06 & 900  & 0.05 & 0.007 & op-thin     \\ \hline
3 & 11.534664 & 42.378428 & M31 & 15.780 & 0.010 & 4.227 & 3715 & 1.53E-07 & 1200 & 0.1  & 0.011 & carbon   \\ \hline
4 & 11.559090 & 42.317375 & M31 & 17.179 & 0.028 & 3.780 & 4078 & 4.04E-06 & 750  & 0.3  & 0.015 & silicate \\
\enddata
\tablecomments{(This table is available in its entirety in machine-readable form.)}
\end{deluxetable*}
\begin{deluxetable*}{ccccccccccccc}
\tablenum{5}
\tablecaption{Catalog of the stellar and dust parameters for red supergiants in M33}\label{cat2}
\tablewidth{0pt}
\tablehead{
\colhead{index} & \colhead{RA} & \colhead{DEC} & \colhead{galaxy} & \colhead{KMag} & \colhead{$\rm KMag\_{Err}$} & \colhead{log$L/L_{\odot}$} & \colhead{$T_{\rm eff}$} & \colhead{ $\dot{M}$ } & \colhead{$T_{\rm in}$} & \colhead{$\tau_V$} & \colhead{$\chi^2$} & \colhead{type}\\
\colhead{} & \colhead{(deg)} & \colhead{(deg)} & \colhead{} & \colhead{(mag)} & \colhead{(mag)} & \colhead{} & \colhead{(K)} & \colhead{ ($M_{\odot}$/yr) } & \colhead{(K)} & \colhead{} & \colhead{} & \colhead{}}
\decimalcolnumbers
\startdata
1733 & 23.222635 & 30.173420 & M33 & 18.085 & 0.081 & 3.502 & 4144 & 9.18E-06 & 750  & 1.05 & 0.015 & silicate \\ \hline
1734 & 23.202302 & 30.172458 & M33 & 16.809 & 0.026 & 3.962 & 3976 & 2.64E-05 & 750  & 1.15 & 0.011 & silicate \\ \hline
1735 & 23.643000 & 30.440888 & M33 & 13.636 & 0.010 & 5.057 & 3405 & 1.04E-04 & 1200 & 1.3  & 0.016 & silicate \\ \hline
1736 & 23.660640 & 30.439871 & M33 & 17.223 & 0.028 & 3.835 & 4108 & 2.44E-05 & 750  & 1.35 & 0.015 & silicate \\ \hline
1737 & 23.551408 & 30.436721 & M33 & 16.008 & 0.010 & 4.246 & 3861 & 3.49E-07 & 1050 & 0.15 & 0.009 & carbon  \\
\enddata
\tablecomments{(This table is available in its entirety in machine-readable form.)}
\end{deluxetable*}

\begin{deluxetable*}{ccccccc}
\tablenum{6}
\tablecaption{The effect of metallicity and surface gravity on the mass loss rate \label{z_logg}}
\tablewidth{10pt}
\tablehead{
\colhead{} & \colhead{[Z]} & \colhead{$\log g$} & \colhead{$\Sigma{\chi^2}$} & \colhead{mean $\dot{M}$} & \colhead{median $\dot{M}$} & \colhead{optically-thin sources} \\
\colhead{} & \colhead{} & \colhead{} & \colhead{} & \colhead{(M$_{\odot}$/yr)} & \colhead{(M$_{\odot}$/yr)} & \colhead{Number}}

\startdata
         This Model& 0 & 0 & 93.9  & 2.01E-5 & 3.08E-6& 1466\\ \hline
        $\log g$=-0.5   & 0 & -0.5 & 101.4  & 2.03E-5 & 3.14E-6& 1436\\ \hline
        $[Z]=0.25$  & 0.25 & 0  & 102.7  & 2.03E-5 & 3.27E-6& 1375\\
\enddata

\end{deluxetable*}
\end{CJK*}
\end{document}